\documentclass[fleqn,10pt]{manuscript}
\usepackage{soul}
\usepackage{graphicx}%
\usepackage{url}
\usepackage{multirow}%
\usepackage{amsmath,amssymb,amsfonts}%
\usepackage{amsthm}%
\usepackage{mathrsfs}%
\usepackage[title]{appendix}%
\usepackage{xcolor}%
\usepackage{textcomp}%
\usepackage{manyfoot}%
\usepackage{booktabs}%
\usepackage{algorithm}%
\usepackage{algorithmicx}%
\usepackage{algpseudocode}%
\usepackage{natbib}
\usepackage{rotating}
\usepackage{tabularx}
\usepackage{threeparttable}
\usepackage{pdflscape}
\usepackage[T1]{fontenc}
\usepackage[utf8]{inputenc}

\title{Solid Target production for Astrophysical Reasearch: the European target laboratory partnership in ChETEC-INFRA}

\author[1,2]{RobertaSpartà}
\author[3]{Alexandra Spiridon}
\author[4,5]{Rosanna Depalo}
\author[6,7]{Denise Piatti}
\author[2]{Antonio Massara}
\author[3]{Nicoleta Florea}
\author[8]{Marcel Heine}
\author[3]{Radu-Florin Andrei}
\author[9]{Beyhan Bastin}
\author[3]{Ion Burducea}
\author[6,7]{Antonio Caciolli}
\author[10]{Matteo Campostrini}
\author[8,11]{Sandrine Courtin}
\author[12]{Federico Ferraro}
\author[2]{Giovanni Luca Guardo}
\author[13]{Felix Heim}
\author[3]{Decebal Iancu}
\author[2]{Marco La Cognata}
\author[14,2,15]{Livio Lamia}
\author[1,2]{Gaetano Lanzalone}
\author[16]{Eliana Masha}
\author[3]{Paul Mereuta}
\author[8]{Jean Nippert}
\author[14,2]{Rosario Gianluca Pizzone}
\author[14,2]{Giuseppe Gabriele Rapisarda}
\author[14,2]{Maria Letizia Sergi}
\author[6,10]{Jakub Skowronski}
\author[3]{Dana State}
\author[17]{Tamás Szücs}
\author[3]{Livius Trache}
\author[1,2]{Aurora Tumino}

\affil[1]{Dipartimento di Ingegneria e Architettura, Università di Enna 'Kore', Enna, Italy}

\affil[2]{Laboratori Nazionali del Sud, INFN, Catania, Italy}

\affil[3]{National Institute of Physics and Nuclear Engineering Horia Hulubei, Bucharest-Magurele, Romania}

\affil[4]{Università degli Studi di Milano, Milano, Italy}

\affil[5]{Sezione di Milano, INFN, Milano, Italy}

\affil[6]{Università degli Studi di Padova, Padova, Italy}

\affil[7]{Sezione di Padova, INFN, Padova, Italy}

\affil[8]{Université de Strasbourg, CNRS, IPHC UMR 7178, 67000, Strasbourg,, France}

\affil[9]{Grand Accélérateur National d’Ions Lourds, GANIL, Caen, France}

\affil[10]{Laboratori Nazionali di Legnaro, INFN, Legnaro, Italy}

\affil[11]{University of Strasbourg Institute of Advanced Studies, USIAS, Strasbourg, France}

\affil[12]{Laboratori Nazionali del Gran Sasso, INFN, Assergi, Italy}

\affil[13]{Institute for Nuclear Physics, University of Cologne, Cologne, Germany}

\affil[14]{Dipartimento di Fisica e Astronomia "E. Majorana", Università of Catania, Catania, Italy}

\affil[15]{Centro Siciliano di Fisica Nucleare e Struttura della Materia, CSFNSM, Catania, Italy}

\affil[16]{Helmholtz-Zentrum Dresden-Rossendorf, Dresden, Germany}

\affil[17]{Institute for Nuclear Research, ATOMKI, Debrecen, Hungary}

\corrauthor[1]{Roberta Spartà}{rsparta@lns.infn.it}

\begin{abstract}
\hl{The joint work of European target laboratories in the ChETEC-INFRA project is presented, to face the new experimental challenges of nuclear astrophysics. In particular, results are presented on innovative targets of $^{12,13}$C, $^{16}$O, and $^{19}$F that were produced, characterized, and, in some cases, tested under beam irradiation. STAR (Solid Targets for Astrophysics Research) is already acting to increase collaboration among laboratories, to achieve shared protocols for target production, and to offer a characterization service to the entire nuclear astrophysics community.}
\end{abstract}

\begin{document}

\flushbottom
\maketitle
\thispagestyle{empty}

\section*{Introduction}
\label{intro}

Reliable and precise nuclear reaction cross section measurements are crucial challenges for nuclear astrophysics. At energies of interest for stellar nucleosynthesis, the cross sections are, indeed, of the order of $pb$ to $fb$, translating to low experimental yields. Scientist efforts are focused on achieving the required precision needed for stellar models via both direct and indirect methods. 

ChETEC-INFRA (Chemical Elements as Tracers of the Evolution of the Cosmos - Infrastructures for Nuclear Astrophysics - https://www.chetec-infra.eu/) %\citet{infrasite} 
is a European project in the framework of H2020 which networks different types of infrastructure, joining forces and expertise needed to study the origin of the chemical elements. 
%Nuclear astrophysics laboratories are among these infrastructures, to supply reactions data. 

\begin{figure}[ht]
\begin{center}
\includegraphics [width=0.6\textwidth] {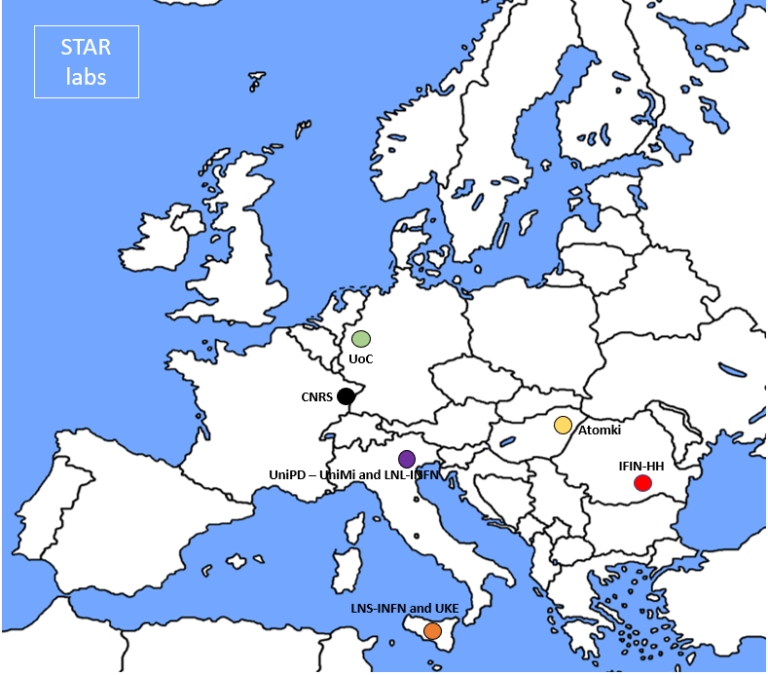}
\caption{Map of the ChETEC-INFRA STAR task members.}
\label{members}
\end{center}
\end{figure}

Nuclear astrophysicists who want to study charged particles induced reactions struggle with the necessity to maximize the signal-to-noise ratio in every possible way, including the target improvement. The development of stable targets for high precision measurements down at astrophysical energies is the main goal of STAR (Solid Targets for Astrophysics Research https://www.chetec-infra.eu/jra/star/) task inside ChETEC-INFRA. %\citet{STARsite}
Besides the TransNational Access facilities (Felsenkeller laboratory, Bellotti-Ion Beam facility, IFIN-HH (Romania)), the STAR participating institutions are: ATOMKI (Hungary), CNRS (France), IFIN-HH (Romania), INFN (Italy), University of Padua, University of Milan, University Kore (Italy), and University of Cologne (Germany), as shown in Figure \ref{members}.

The STAR collaboration aims to develop, test and make protocols of special solid targets complying with the new requirements of experimental studies of reactions of astrophysical interest, namely purity, enrichment, composition, thickness, and stability under irradiation. 

To this end, STAR is working towards the realization of:

- ultra-pure material targets for low reaction yields, in which signals from parasitic reactions on impurities can limit experiments and must therefore be reduced.

- noble gases solid targets: being impossible for them to create solid compounds, noble elements are generally used in the form of gas target, but other than their low densities, angular spread and energy loss and straggling can hinder the reaching of the experimental needs in many cases. Thus, He and Ne solid targets, produced by implantation into a host material, would be preferable.

STAR activities are mainly focused on targets suitable for crucial measurements for the astrophysical scenarios of C-burning (sec. \ref{res}), s-process (sec. \ref{s-proc}) and of physics beyond the Standard Model (sec. \ref{beyond}).

C-burning main process is the $^{12}$C+$^{12}$C fusion reaction (sec. \ref{lnl}), which plays a key role in the heavy elements nucleosynthesis and for the supernovae (SNe) evolution, significantly affecting the chemical and physical galaxy evolution. At astrophysical energies, $^{12}$C+$^{12}$C fusion reaction cross sections are of the order of $fb$. Present experimental direct data (lowest E$_{c.m.}$=2.2 MeV) are affected by large uncertainties, that are further enhanced when extrapolated to stellar energies, because of the poorly constrained structure of the $^{24}$Mg compound nucleus \citet{Fruet2020, Tan2020, Jiang2018, Spillane2007}. A recent indirect measurement \citet{Tumino2018} down to low energy has found a resonant behaviour which is also responsible of a huge increase in the reaction rate in the temperature range of interest for superburst (T=0.4-0.5 GK).

To substantiate these results and to put on a firmer ground the normalization of the indirect result, a future measurement is planned by the LUNA collaboration, for which STAR collaborates in the search for the optimal target. 

%been questioned \citet{Mukh2020} 

Moreover, also other light ions fusion reactions, as $^{12,13}$C+$^{16}$O or +$^{19}$F, are under consideration by researchers to investigate the presence of specific reaction mechanisms like the so-called \textit{fusion hindrance} at sub-barrier energies (sec. \ref{13c16o} and \ref{13c19f}).  

Highly enriched fluorine targets were produced and tested both for the latter and for the X17 boson \citet{Krasznahorkay} search in the NEW JEDI experiment \citet{Beyhan} through the study of $^{8}$Be levels in the 18 MeV energy region, populated via $^7$Li+p. The proton beam was impinging on a LiF target, hence the need of fluorine targets for the background subtraction. Results are in sec. \ref{beyond}.

The nucleosynthesis via s-process is responsible for the production of about half of heavy elements with A$\,\le\,$90. The two main neutron sources are the $^{13}$C($\alpha$, n)$^{16}$O and the $^{22}$Ne($\alpha$,n)$^{25}$Mg reactions. While the former has been recently measured inside the Gamow Peak \citet{Ciani2021,Gao2022,Laco} the latter is still poorly constrained below the resonance at E$\sim$832 keV (J$^{\pi}$ = 2$^+$) \citet{Jaeger,Harms91,Drotleff1993}, where mainly upper limits are reported. Moreover, a crucial parameter to improve our understanding of s-process nucleosynthesis is the ratio between ($\alpha$,n) and ($\alpha$,$\gamma$) rates, since it determines the temperature at which the neutron production starts to become effective \citet{ads21}.
%While LUNA collaboration has been measuring both the $^{22}$Ne($\alpha$,n)$^{25}$Mg and the $^{22}$Ne($\alpha$,$\gamma$)$^{26}$Mg reaction cross sections at the Bellotti Ion Beam Facility using an extended gas target, 
STAR studies for solid helium targets for future measurement planned at INFN-LNS are explained in sec. \ref{s-proc}. 

It is worth to promote that target production and characterization is a service open to the nuclear astrophysics European community (for contacts, see \citet{STARsite}).

Results achieved in the STAR framework are described in the following sections.

\section{C-burning and light ions sub-barrier fusion hindrance of the cross sections} \label{res}

The study of fusion reactions occurring at energies below the Coulomb barrier between light ions, especially $^{12}$C and $^{16}$O (thus $^{12}$C+$^{12}$C, $^{12}$C+$^{16}$O, $^{16}$O+$^{16}$O), holds a large significance in various astrophysical scenarios, spanning from the creation of elements to the occurrences of neutron star superbursts resulting from accretion. 
Experimental challenges at these energies obstruct even a gross knowledge of these cross sections, thus also the possibility of a strong fusion hindrance was considered. 
This idea comes from the extension to light nuclei fusion of the low energy fusion cross section hindrance of medium and high mass ions, which is now established for these negative Q-value fusion reactions \citet{Misicu,Jiang2008}. For lighter ion-ion reactions hindrance causes assumed are the incompressibility of nuclear matter and the Pauli principle.
For these reasons, studies of the systematics of these reactions and reactions between nearby nuclei at sub-Coulomb energies are still undergoing. $^{12}$C + $^{12}$C is one of the most important reactions in nuclear astrophysics, due to the strong influence that carbon burning has on the fate of massive stars \citet{chieffi} and super-bursts from accreting neutron stars \citet{schatz}. 

According to present understanding, carbon burning occurs at temperatures of about 800 MK, corresponding to the reactions effectively occurring in the energy range E$_{c.m.}$ = 1.0 - 2.0 MeV, with sub-$fb$ cross sections.
Direct measurements, so far, reported results affected by high uncertainty down to about E$_{c.m.}$ = 2.2 MeV \citet{Becker1981, BARRONPALOS2006, Spillane2007, Jiang2018, Zickefoose2018, Fruet2020, Tan2020}, with resonances adding extra uncertainties. Besides the extremely low cross section, environmental and beam induced background was the main limitation for direct measurements, the latter due to H, D and $^{13}$C contaminations in carbon targets \citet{Becker1981, BARRONPALOS2006, Spillane2007, Jiang2018, Zickefoose2018, Fruet2020, Tan2020}. 

Multiple experimental efforts are focused on performing direct measurements of carbon fusion cross section and measurements with improved accuracy right inside the astrophysics region of interest of core carbon burning in heavy stars \citet{Fruet2020, Tan2020, tan2024, monpribat2022}. The measurement setups are designed for reliable and enduring detection of the charged particles (alphas and protons) evaporated from the compound nucleus $^{24}$Mg in coincidence with gammas from the de-excitation of the daughter nuclei ($^{20}$Ne, $^{23}$Na) with efficient background suppression for the accurate determination of cross sections of 100~$nb$ or below \citet{Heine,heine2022EPJWeb}. This necessitates measurements campaigns of typical several weeks with bombardment of the carbon targets with carbon beam of an intensity of up to 10~p$\mu$A. Under such high-duty conditions, accurate knowledge of the intact state of the target are indispensable~\citet{aguilera2006NIMB, tan2024} and multiple characterization before, during and after beam exposure are combined during the measurements with STELLA (STELlar LAboratory) in order to guarantee accurate measurements \citet{Heine, nippert2024}, as it is shown in \ref{stella}.

Also, LUNA collaboration proposed a new direct measurement at the deep underground Bellotti Ion Beam Facility (B-IBF) of LNGS, Italy. LUNA aims at measuring the cross sections of the $^{12}$C($^{12}$C,$\alpha$)$^{20}$Ne and the $^{12}$C($^{12}$C,p)$^{23}$Na reactions by detecting $\gamma$-rays emitted in the de-excitation of the $^{23}$Na and $^{20}$Ne nuclei, produced in an excited state after the emission of a charged particle by the compound nucleus. This approach would benefit the most from the deep underground location of the B-IBF, where cosmic rays are suppressed by six orders of magnitude, reducing to almost zero the high energy $\gamma$ background and enabling the use of thicker lead shielding around the detector. The dynamic range of LUNA-MV 3.5 Singletron accelerator is 1-7 MeV, allowing to perform a comprehensive measurement of the $^{12}$C$\,+\,^{12}$C reaction cross section. Moreover the expected $^{12}$C$^+$ and $^{12}$C$^{++}$ beam intensities at the B-IBF are up to 150 $\mu$A \citet{Sen2019}, which allows to achieve the proper sensitivity \citet{Piatti2024}, but still the target to be used is a crucial element for these measurements, as detailed in sec. \ref{luna}. 

Besides direct approaches, recently a measurement of the $^{12}$C+$^{12}$C cross section was performed with Trojan Horse Method (THM), which is based on the transfer of a $^{12}$C nucleus to $^{12}$C in the three-body processes $^{12}$C($^{14}$N,$\alpha ^{20}$N)$^{2}$H and$^{12}$C($^{14}$N,p$^{23}$Na)$^{2}$H in the quasi-free kinematic mechanism \citet{Tumino2018}. The extracted astrophysical S-factor presents several resonances that enhances the reaction rate by a factor of up to one hundred, with significant impact on our present knowledge of the carbon burning.

%The analysis of \citet{Tumino2018} has been recently questioned in \citet{Mukh2020}, which substantially reduces the overall cross section by up to three orders of magnitude, causing a reduction in the reaction rate. 
%None of these predictions include the possibility of hindrance \citet{Jiang2007} due to nuclear incomprehensibility of $^{12}$C in the low energy fusion process, which would deserve a direct study.

The THM data do not suggest any hindrance effect and this is also what has been found on $^{12}$C+$^{13}$C fusion by \citet{Zhang}. More specifically, with respect to the ongoing hindrance-debate for low mass fusion reactions \citet{Hind1, Hind2, Hind3}, the results in \citet{Zhang} suggest that a maximum in the S-factor is not in fact present for the $^{12}$C + $^{12}$C reaction. %, an observation confirmed by the above-mentioned Trojan Horse measurement \citet{Tumino2018}.
Indeed, while $^{13}$C + $^{12}$C does not show a resonant behavior as $^{12}$C + $^{12}$C, it was determined that it does indicate an upper limit for the resonant cross sections as well as a general trend for the behavior of the astrophysical S-factor \citet{Notani}. 

Moreover, a complementary study was done by the Nuclear Astrophysics Group (NAG) at “Horia Hulubei” National Institute for Physics and Nuclear Engineering (IFIN-HH) using the local 3 MV Tandetron and the activation method. The reaction was studied through the adjacent system $^{13}$C + $^{12}$C. The proton evaporation channel produces the unstable $^{24}$Na with a half-life of 15 hours and its deactivation could be measured inside the Sl\u{a}nic salt mine in an ultra-low background environment \citet{Dana_NIM}. This allowed for the determination of the fusion cross section at deep sub-barrier energies, with the goal of determining the low energy trend of the fusion cross section and of testing the hindrance hypothesis, that should be observed in all of the exit channels and in adjacent systems.

Similarly, in order to help the current understanding of fusion mechanisms at sub-Coulomb energies in low-mass ion-ion systems, the NAG group has continued to study similar reactions (see subsections \ref{13c16o} and \ref{13c19f}). Because activation was found to be the most sensitive method, the reactions studied were chosen based on the activation channels produced as well as on the combination of beam intensities and targets available. Below we show the various targets that were produced and tested for these studies.

\subsection{$^{12}$C+$^{12}$C} \label{lnl}

\subsubsection{The STELLA rotating target system} \label{stella}

The STELLA collaboration is engaging in direct carbon fusion measurements with an innovative rotating-target mechanism employing large thin self-supporting carbon foils~\citet{Heine, nippert2024}. In this approach, the beam traverses the foil undergoing an energy loss of typically 100~keV, and subsequently its charge is determined in a separate beam dump (Faraday cup) for data normalization. The beam hits an outer trajectory of the target-foil surface (target frame diameter: 6.3~cm), that is spinning with of up to 1000~rpm and effectively cooling down by radiating when rotated out off the nominal beam spot position. During data taking periods, the intact status of the target foil is online monitored by the ratio of the counting rate of the beam dump to elastic beam-scattering observed with dedicated silicon detectors further away from the target.

\begin{figure}[ht]
    \begin{center}
    \includegraphics [width=.5\textwidth] {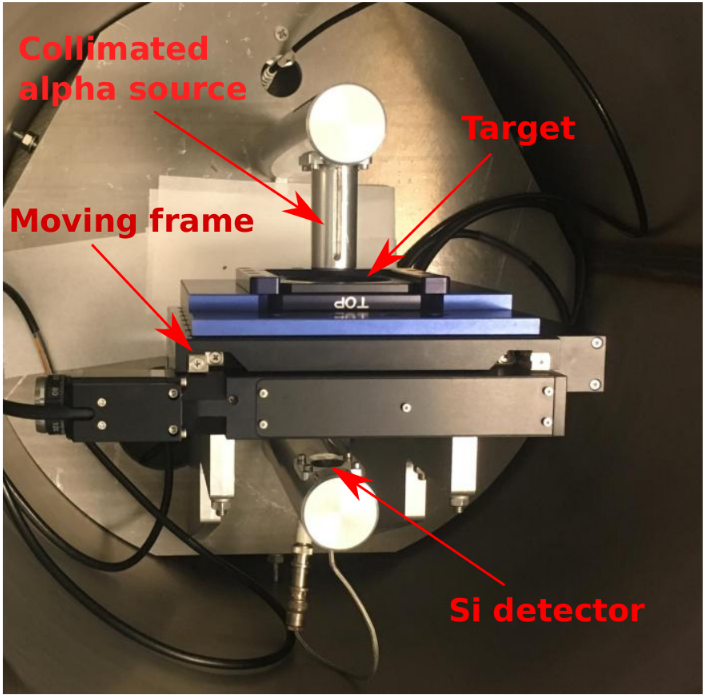}
    \caption{\label{fig:scan_stelYork} Photograph of the target foil measurement station for characterizing large thin carbon target foils at IPHC Strasbourg.}
    \end{center}
\end{figure}

\begin{figure}[ht]
    \begin{center}
    \includegraphics [width=.7\textwidth] {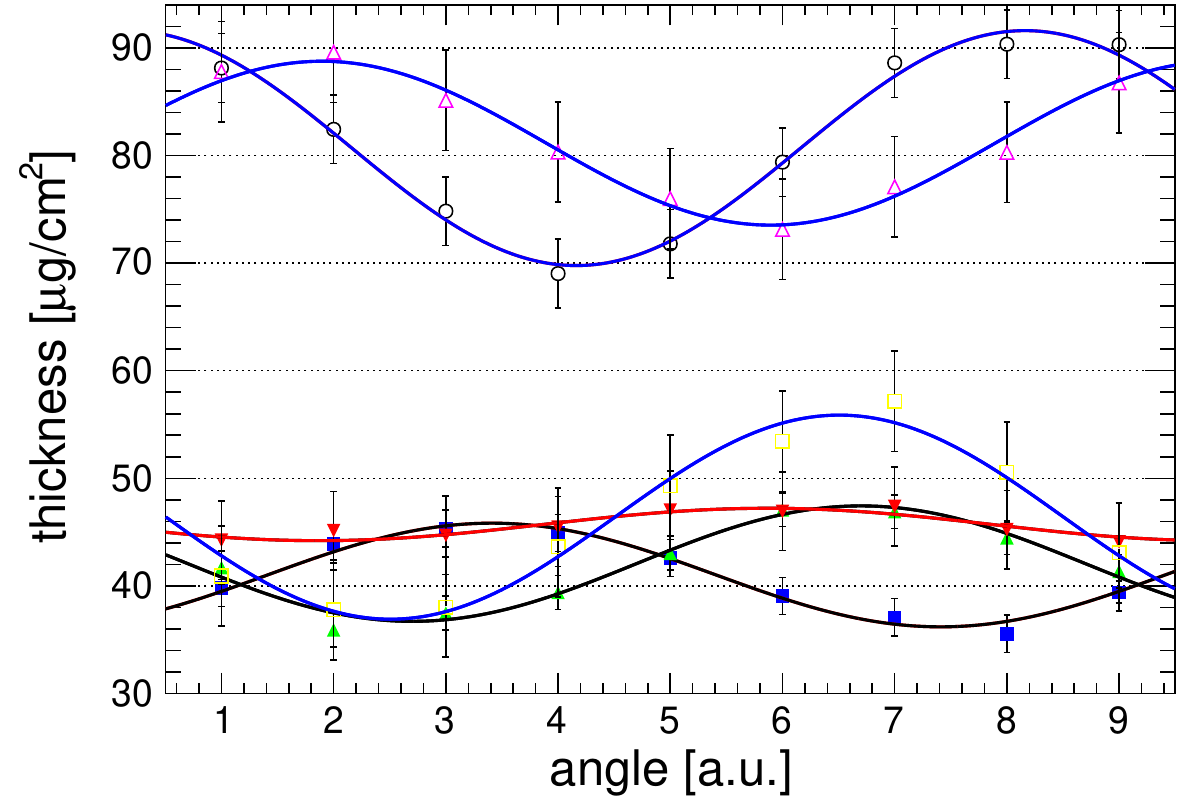}
    \caption{Sequences of target thickness determination measurements at IPHC Strasbourg for large thin carbon targets. Data are taken in iterative measurements on and off the irradiated target area and several target production patches are indicated by the color code of the fitting curves (see text for details).}\label{fig:scan_stella} 
    \end{center} 
\end{figure}

The thickness of the target foils is obtained upon production by high-precision mass determination. These measurements are accompanied by samples of Raman-spectroscopy analysis of the carbon target foil material after beam exposure~\citet{Fruet2020} and systematic thickness determination with an alpha-particle emitting source measuring the thickness dependent energy loss. To this purpose, the measurement station displayed in Figure~\ref{fig:scan_stelYork}
was developed in collaboration with the University of York. The device hosts an alpha emitter, a mobile target holder and a silicon detector. In a series of measurements alternating between points on and off area with beam exposure, the energy loss is determined and converted into the target thickness. Note that the beam spot forms a circular track on the STELLA targets from the rotating system. A measurement hence yields a sequence of eight data points along the outer radius of a target foil with an equal amount of reference points at the center to guarantee the accuracy of the method. A variety of such scans is presented in Figure \ref{fig:scan_stella}

as a function of the angle along the beam track. Data follow a sine curve indicating a slight thickness gradient of the target foils (see 
\citet{nippert2024} for details) and several target-foil patches are grouped by the color code. For the normalization of the $^{12}$C+$^{12}$C measurements, the average target thickness, which is in good agreement with the thickness determination upon target production, is used. This routine improves the precision of the target thickness to better than 10$\%$ and furthermore excludes carbon built-up on the target~\citet{healy1997, aguilera2006NIMB, Fruet2020} demonstrating the reliability and accuracy of the target foil normalization during the $^{12}$C+$^{12}$C measurements with STELLA. Such routines are in high need, seen the challenges of exact target thickness determination in such experiments~\citet{Tan2020, tan2024}.

\subsubsection{Contaminants desorption with HEAT and the tests on carbon targets for the LUNA measurement} \label{luna}
The presence of a strong beam induced background due to the interaction of the carbon beam with hydrogen and deuterium contamination inside the targets: $^1$H($^{12}$C,$\gamma$)$^{13}$N, $^{2}$H($^{12}$C,p$\gamma$)$^{13}$C and the two-step process $^2$H($^{12}$C, $^2$H)$^{12}$C $\rightarrow$ $^{12}$C($^2$H,p$\gamma$)$^{13}$C hamper the cross section measurement at low energies \citet{Zickefoose2018, Morales18, Spillane2007}. 
In literature, irradiating the target for few minutes, heating it up to hundreds of Celsius degree, was reported as an efficient technique to reduce the hydrogen contamination \citet{Spillane2007}. However beam heating lacks of reproducibility. Despite the thick target approach is favored over thin target, which suffer of C-buildup problem \citet{Becker1981}, other technical difficulties arise, such effective cooling of the target and detectors closed by, radiated heat and deposition of carbon inducing structural changes in the target \citet{Spillane2007, Picollo2021}.

\begin{figure}[ht]
\begin{center}
\includegraphics [width=\textwidth] {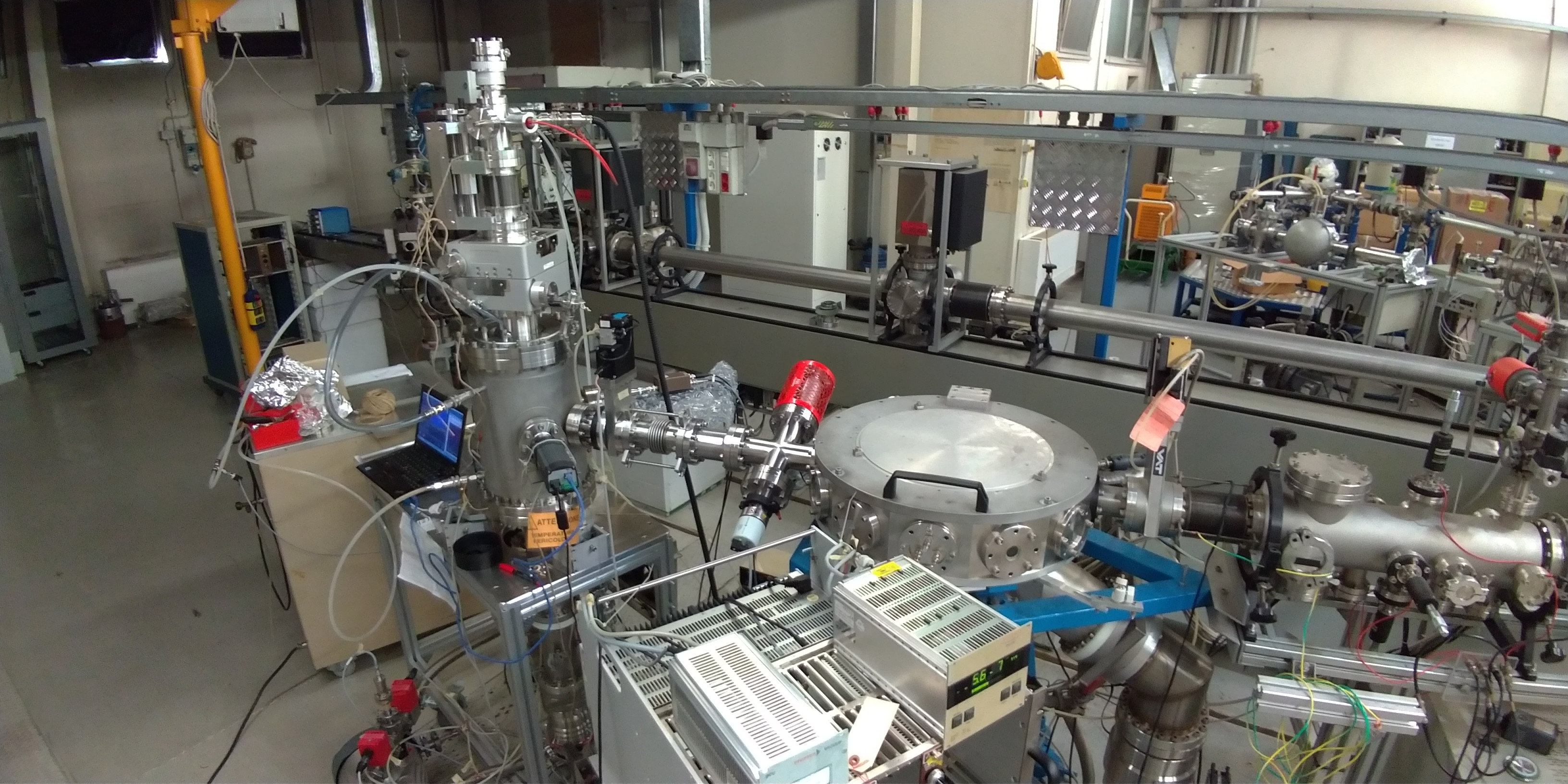} \\
\caption{Picture of the HEAT setup installed at the AN2000 accelerator facility of Legnaro National Laboratories.}
\label{fig:heat}
\end{center}
\end{figure}

The HEAT project (Figure \ref{fig:heat}) at the Laboratori Nazionali in Legnaro (LNL) of INFN, in Italy, aims at developing pure carbon targets for measurements of the cross section of the $^{12}$C+$^{12}$C fusion reaction. The goal of the HEAT experiment is to establish a reproducible technique for hydrogen desorption from different types of carbon targets, to ideally decrease the hydrogen content at the level of a few ppm. 

\begin{figure}[ht]
\begin{center}
\includegraphics [width=0.85\textwidth] {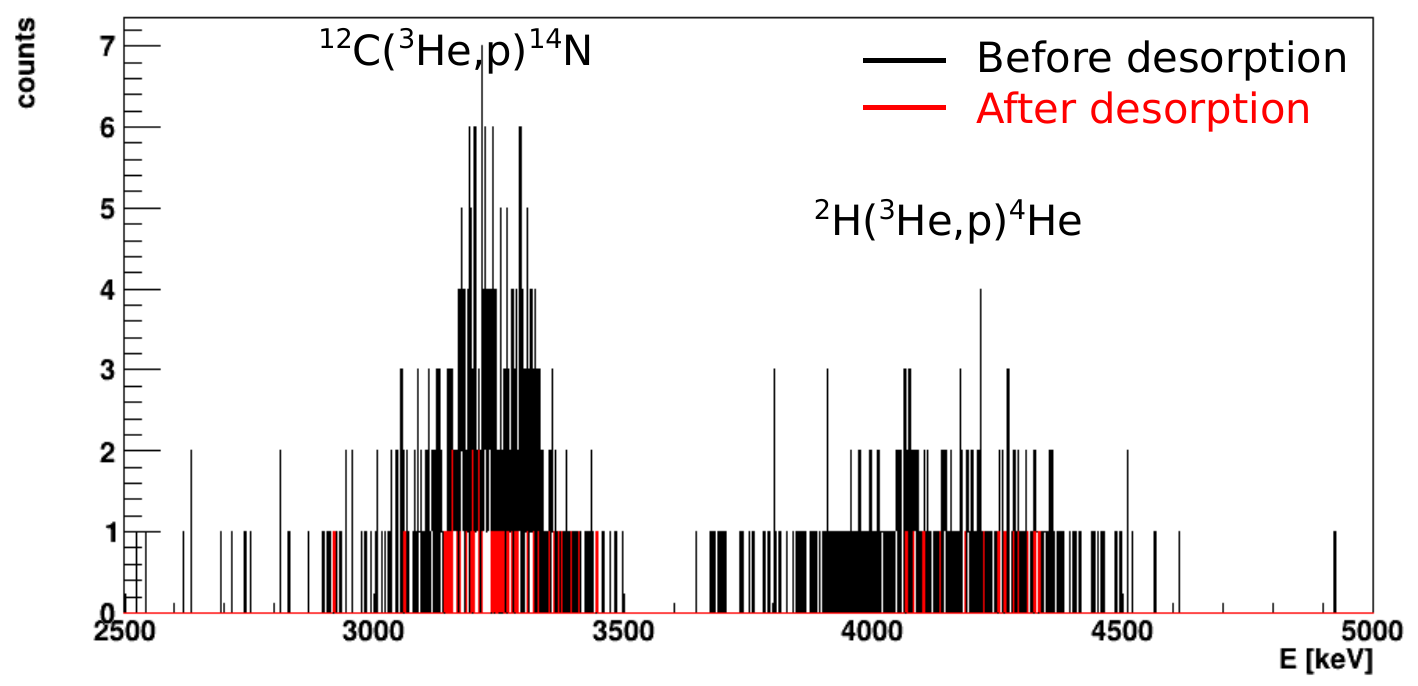} \\
\caption{Natural graphite (0.95\% nominal purity) spectra before and after a 1 MeV beam of $^3$He.}
\label{fig:heatspectra}
\end{center}
\end{figure}

Targets investigated so far include natural graphite, HOPG and glassy carbon \citet{depalo21}. The measurements confirm that HOPG is the graphite with the lowest content of hydrogen. In natural graphite, desorption levels as high as a factor of 3 have been observed (see Figure \ref{fig:heatspectra}). Desorption tests on tantalum, one of the most common backing material for solid targets, have also been performed, and the setup could be used in the future for different applications. 

\begin{figure}[!ht]
\begin{center}
\includegraphics [width=0.85\textwidth] {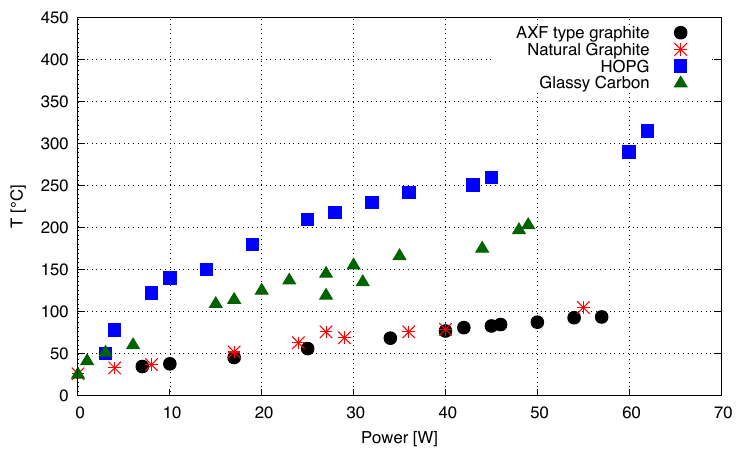} \\
\caption{AXF ENTEGRIS graphite (black circles), Natural graphite (red stars), HOPG (blue squares) and glassy carbon (green triangles) temperature of the beam spot. The ENTEGRIS graphite shows the best thermal conductivity and after long irradiation (up to 3 C) showed the best endurance. HOPG and Glassy C showed, indeed, a swelling beamspot and a crack respectively.}
\label{carb}
\end{center}
\end{figure}

HEAT studies for the amount of H and D contaminants were found to be relevant for the target choice of the $^{12}$C+$^{12}$C cross section measurement planned by LUNA collaboration at the B-IBF at LNGS. 
Additionally, LUNA collaboration tested the different types of target stability under irradiation and their thermal behavior for a properly designed target holder. The first tests have been performed at the shallow underground facility of Felsenkeller. A 8.8 MeV beam of $^{12}$C$^{+++}$ , with intensity 2-5 p$\mu$A, was delivered to natural graphite, HOPG, glassy carbon and AXF and ZXF-type graphite from ENTEGRIS, as shown in Figure \ref{carb}. 
These targets were selected because of either low H and H$_2$ content (HOPG), good thermal resistance (Glassy carbon and graphite) or conductivity (HOPG) and their nanostructure (AXF and ZXF-type graphite). From preliminary analysis the AXF and ZXF graphite proved to be the most stable in terms of thermal and structural behavior, and for this reason, they are good candidates for the LUNA measurement. Analysis is ongoing to check the contamination level in each different target as well as the beam induced degradation in each of them. 
%Moreover, new tests with higher intensity $^{12}$C$^{+}$ beam are planned in 2024 at the B-IBF. 
%As an output of both tests a technical paper will come out and the optimal target will be used for the planned measurement by LUNA in 2025.

\subsection{$^{12,13}$C+$^{16}$O} \label{13c16o}

As mentioned at the beginning of this section, fusion reactions between light ions, especially $^{12}$C and $^{16}$O, play an important role in stellar evolution and explosive nucleosynthesis. The reaction $^{12}$C+$^{16}$O has been found to have a significant contribution in both the carbon and oxygen burning stages of stars \citet{Wiescher2012,Woosley1971}. While $^{12}$C+$^{12}$C is the main reaction during core and shell carbon burning, towards the end of this phase as $^{12}$C nuclei are depleted, the abundance of $^{16}$O becomes substantially higher as a result of the $^{12}$C($\alpha$,$\gamma$)$^{16}$O \citet{Weaver1993,Buchmann1996}. A similar scenario occurs in oxygen burning, where the $^{16}$O+$^{16}$O reaction is dominant. Here, as $^{16}$O nuclei are depleted, the photodissociation of $^{16}$O increases the abundance of $^{12}$C nuclei leading to potential reaction with $^{16}$O. In both cases, the competitiveness of the $^{12}$C+$^{16}$O is enabled by the temperature and density increases, characteristic of late stage burning. It has also been shown in recent studies \citet{MR2017ApJ} that this reaction can play an important role in the evolution of Type Ia supernovae, depending on the relevant fusion rates and environmental conditions.
% As shown by recent studies [6], the 12 C + 16 O fusion can also play a significant role in this scenario, at temperatures around 3.6 · 10 9 K. Therefore, to assess the impact of the 12 C + 16 O fusion in the whole context of carbon burning, a precise study of its reaction rate in the energy range of astrophysical interest, namely between 3 and 7.2 MeV in the center of mass frame, is required.

% Status of the 12 C + 16 O
It is, therefore, similarly important to perform studies of the sub-Coulomb barrier fusion cross section of the $^{12}$C+$^{16}$O reaction. So far, in literature, most of the measurements that can be found for this reaction were performed in the 1970s \citet{Patterson1971,Christensen1977,Cujec1976} with oxygen beams and natural carbon targets. However, they all stop at a center of mass energy of $\sim$4 MeV leading to the necessity of using extrapolation methods for lower energies. A more recent study was performed by Fang et al. \citet{Fang2017} using single and particle-$\gamma$ coincidence techniques, which extended the known experimental range down to $\sim$3.6 MeV and hinted at the presence of the hindrance phenomenon. Even more recently, in \citet{Oliva2023EPJWC} is reported the completion of an experiment at the Laboratori Nazionali del Sud, Italy, that used the indirect THM method to evaluate the $^{12}$C($^{16}$O,$\alpha$)$^{24}$Mg and $^{12}$C($^{16}$O,p)$^{27}$Al reactions at astrophysical energies, but data analysis is still ongoing.
%  Unfortunately, this procedure is strongly dependent on the model used, on the presence of nuclear effects not included in the model (clustering, molecular-like behaviour, etc.) and on the stellar environment itself [8]. Therefore, a new measurement is needed to better explore the low-energy region. During both hydrostatic and explosive carbon burning, the 12 C + 16 O fusion process proceeds mainly through 12C( 16 O, α) 24 Mg (Q = 6.77 MeV) and 12 C( 16 O, p) 27 Al (Q = 5.17 MeV) reaction channels and partially through the 12 C( 16 O, n) 27 Si, less favored due to its negative Q-value. Other reactions such as 12 C( 16 O, 2α) 20 Ne are hindered by the presence of the Coulombian barrier in the exit channel [7].

%%%%%%%%%%%%%%%%%%%%%%%%%%%%%%%%%%%%%%%%%%%%%%%%%%%%%%%%%%%%%%%%%%%%%%%%%%%%%%%%%%%%%%%
Because of the many open questions left by other studies, at IFIN-HH, the $^{12}$C + $^{16}$O reaction was studied through the neighboring reaction, $^{13}$C + $^{16}$O, using the very sensitive activation method in order to determine partial cross sections, and therefore the total fusion cross section. This reaction produces three radioactive isotopes, $^{28}$Al (T$_{1/2}$ = 9.8 min), $^{27}$Mg (T$_{1/2}$ = 2.24 min) and $^{24}$Na (T$_{1/2}$ = 15 hours), and as such the focus was on measuring gamma decay spectra down to the lowest energies possible. Experimentally, this was approached in two ways: a $^{16}$O beam on $^{13}$C targets and a $^{13}$C beam on $^{16}$O targets.
%For this purpose, two thin $^{13}$C targets were produced and tested, but it turned out that they do not resist under intense beams. Therefore, $^{16}$O targets as cerium-oxide (CeO$_2$) and tantalum pentoxide (Ta$_2$O$_5$) were prepared and tested, where they resisted in-beam.

\subsubsection{$^{13}$C targets}

\begin{figure}[ht]
\begin{minipage}[t]{6.5cm}
\centering
\includegraphics[width=0.7\textwidth]{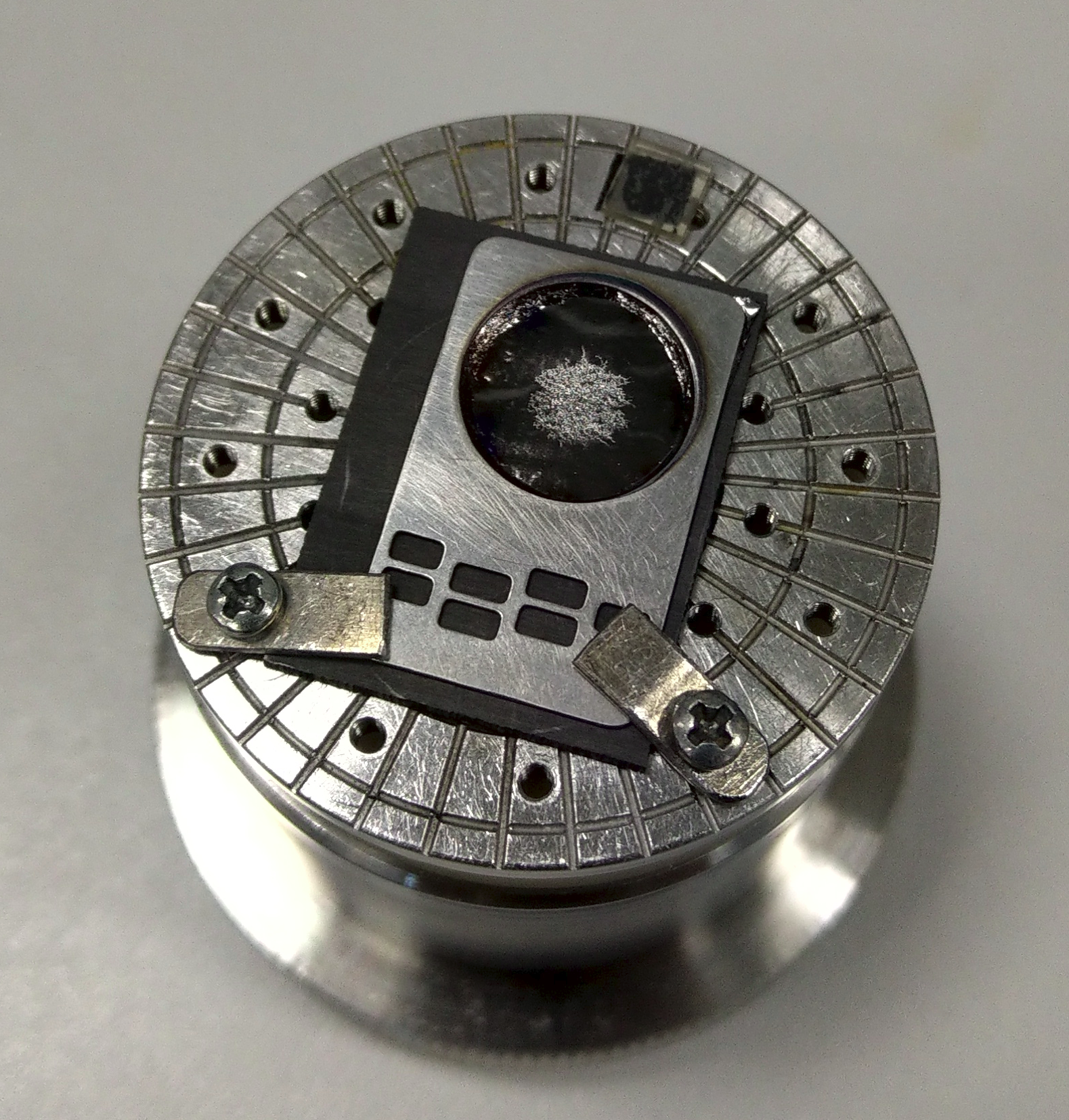}
\caption{Picture of a $^{13}$C targets made by deposition on Ta backing.}\label{fig:13Ctgt}	
\end{minipage}
\ \hspace{2mm} \hspace{2mm} \
\begin{minipage}[t]{6.5cm}
	\centering
	\includegraphics[width=0.95\textwidth]{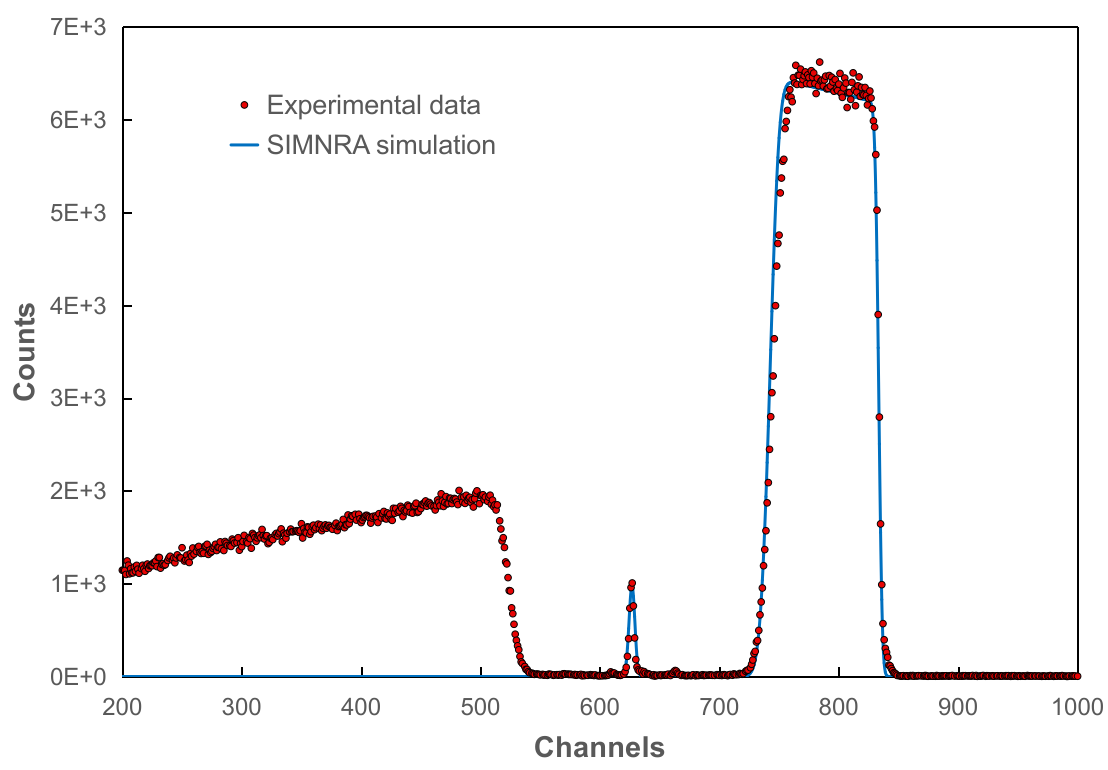}
	\caption{RBS spectrum for the target on the left, showing the small peak corresponding to $^{13}$C and the larger one denoting the Ta backing. The slight disagreement between the simulation (blue line) and the experimental data (red dots) is due to the presence of rugosities on the surface (as can also be seen in Figure \ref{fig:13Ctgt}).}
	\label{fig:13Ctgt_RBS}
\end{minipage}
\end{figure}

The first approach, as mentioned above, was to use an $^{16}$O beam on a $^{13}$C target. The oxygen beam was produced from NiO at energies E$_{lab}$=10-14.8 MeV with currents between 3 p$\mu$A and 240 pnA, respectively. Thin $^{13}$C targets were prepared by deposition of $^{13}$C (from powder with 99\% purity) on Ta backing (a foil of 3 mg/cm$^{2}$ thickness) using the electron-gun method part of the TE18-High Vacuum Deposition System produced by Intercovamex company. Figure \ref{fig:13Ctgt} shows a picture of one such target. 
The thickness of the deposited $^{13}$C layer for each target was determined using the RBS technique, using a proton beam from the 3 MV Tandetron at IFIN-HH accelerator at energy of 2.5 MeV. An example of RBS spectrum is shown in Figure \ref{fig:13Ctgt_RBS}, showing the data (red circles) for the central spot of the target in Figure \ref{fig:13Ctgt}, as well as the simulation performed with SIMNRA 7.0 \citet{SIMNRA} (blue line). The thickness of these targets was determined to be, on average, 170 nm (40 $\mu$g/cm$^{2}$). A visual assessment of the targets showed the presence of significant rugosities in the beam spot area, an effect that leads to thickness non-uniformity and uncertainties in cross section estimations. 
%figure5

\begin{figure}[ht]
\begin{minipage}[t]{5.5cm}
\centering
        \includegraphics[width=0.6\textwidth]{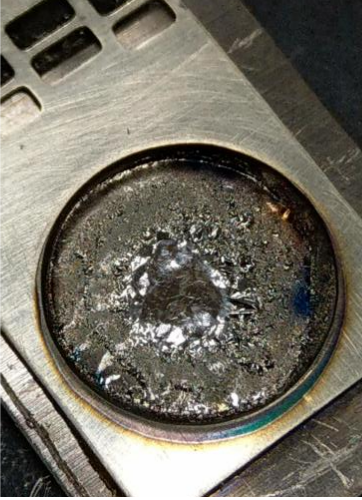}
	\caption{Picture of one $^{13}$C target after irradiation with beam currents of $\sim$1 p$\mu$A.}
	\label{fig:ifin-6}
\end{minipage}
\ \hspace{8mm} \hspace{4mm} \
\begin{minipage}[t]{6.5cm}
	\centering
	\includegraphics[width=0.95\textwidth]{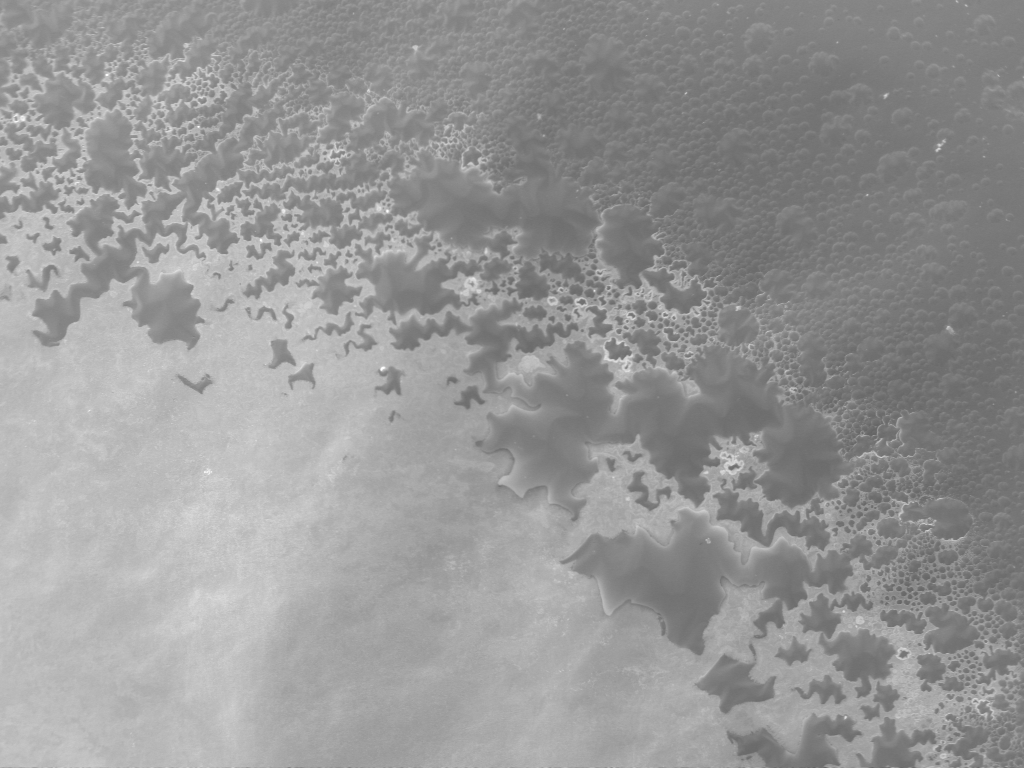}
	\caption{Picture taken of target in Figure \ref{fig:ifin-6} with SEM at 100x magnification, showing the transition between the beam spot (polished surface) and the rest of the target surface (rough surface).}
	\label{fig:ifin-7}
\end{minipage}
\end{figure}

Given the short half-lives of the unstable nuclei produced ($^{27}$Mg with T$_{1/2}$=9.8 min and $^{28}$Al with T$_{1/2}$=2.2 min), the targets were used for repeated measurements of 30 - 60 minutes irradiation followed by 50 minutes decay measurements. The targets were assessed on a visual level after each irradiation and found to be deteriorating even when lower beam currents were used. Specifically, it was determined that the heat generated by the beam was causing the carbon layer to form bubbles, eventually leading to complete detachment (Figure \ref{fig:ifin-6}). %(exfoliation?) and loss of material in the beam spot (Figure \ref{fig:ifin-6}). %It's also about the fact that we can't estimate when the layer detached completely so we can't correct our data for those uncertainties.
%Is it worth it to add a gamma spectrum somewhere here? Maybe one that can show the problem above.
%figures67

It should be noted that the flange holding the target was cooled with a water-based system during irradiation, but it appears that it was insufficient. Post-irradiation Scanning Electron Microscopy (SEM) confirmed the absence of carbon in the beam spot (Figure \ref{fig:ifin-7}), leaving  only the Ta backing with flakes of $^{13}$C on the border.

Given these test results and that the NAG group's interest is in very low energies and low cross sections measurements, where high beam currents are paramount, it was decided that such thin targets could not be used for these purposes.\\

\subsubsection{Oxygen targets}

The second approach was to use a $^{13}$C beam on an $^{16}$O target. The carbon beam was produced at different laboratory energies between 8 and 15 MeV with currents between 3 p$\mu$A and 200 pnA, respectively. Several oxide compounds were tested, chosen primarily based on two criteria: the Z of the compound partner (higher means higher Coulomb barrier, therefore a lower probability to induce a contaminant reaction with the beam) and the number of oxygen atoms in the compound.

\paragraph{(a) Cerium dioxide (CeO$_2$)}

The first compound that was tested was cerium (IV) oxide. Targets were prepared from CeO$_2$ powder using the tablet pressing method.
%and the Specac Atlas Hydraulic Press Automatic 25 Ton.  
The powder was pressed into pellets by applying a load of 25 ton.
%(where T is the unit of measurement of the hydraulic presses).
The diameter of the pellets was set at 2 cm or less in order to fit inside the target holder, but different thicknesses were attempted: 0.8 g/cm$^2$ and 1 g/cm$^2$. It was found that the targets could not be made any thinner than 0.75 g/cm$^2$, as they became too fragile to handle and crumbled immediately.
%figure8 
%\begin{figure}[ht]
%\begin{center}
%   \includegraphics[width=0.55\textwidth]{Ifin-CeO2_new.png}
%	\caption{CeO$_2$ solid targets.}
%	\label{fig:ifin-8}
%\end{center}
%\end{figure}

\begin{figure}
\begin{minipage}[b]{6.5cm}
\centering
	\includegraphics[width=0.55\textwidth]{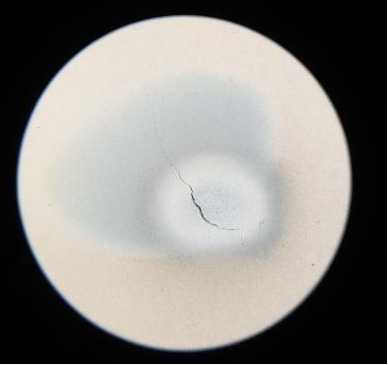}
	\caption{CeO$_2$ target seen with microscope at 8x magnification showing damage after 1 mC irradiation.}
	\label{fig:ifin-CeO2_T1}
\end{minipage}
\ \hspace{2mm} \hspace{2mm} \
\begin{minipage}[b]{6.5cm}
\centering
    \includegraphics[width=0.55\textwidth]{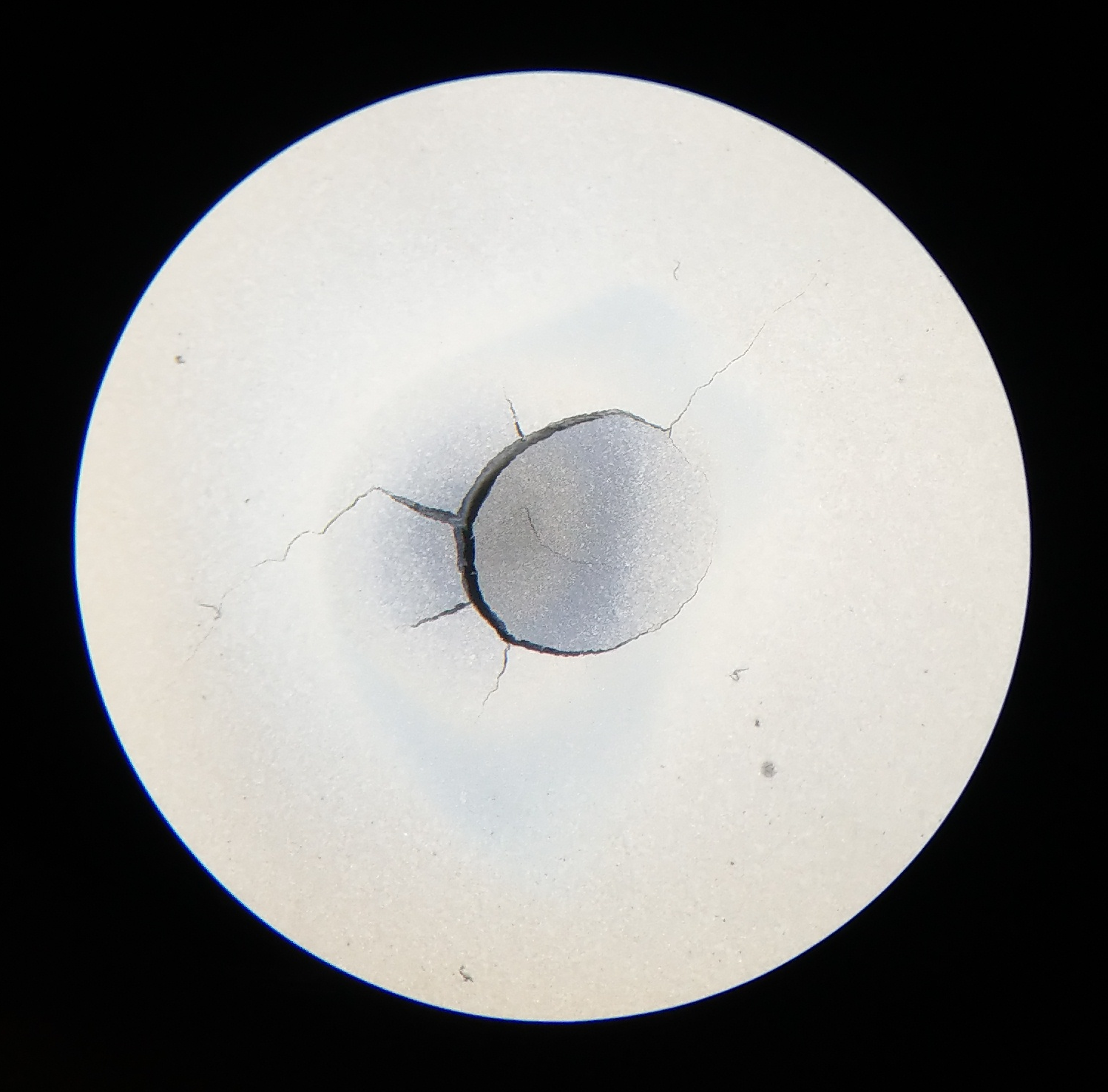}
	\caption{CeO$_2$ target seen with microscope at 8x magnification showing damage after 10 mC irradiation.}
	\label{fig:ifin-CeO2_T7}
\end{minipage}
\end{figure}

In order to provide support and easier handling, the pellets were attached to aluminum backing in the shape of discs ($\sim$1 mm thick). The targets were irradiated on average for 60 minutes and were affected on various levels by the beam, depending on the current. Figure \ref{fig:ifin-CeO2_T1} shows the beam spot of a CeO$_2$ target (0.8 g/cm$^2$) after 40 min of irradiation with $^{13}$C at 10.8 MeV and a current of $\sim$80 pnA ($\sim$1 mC beam charge). Similarly, Figure \ref{fig:ifin-CeO2_T7} shows a different target (0.8 g/cm$^2$) after a slightly longer irradiation (60 minutes) but with a beam current of $\sim$1 p$\mu$A ($\sim$10 mC beam charge).
%figures910  change this pic maybe or adjust it more

\begin{figure} [!ht]
\begin{minipage}[t]{7.5cm}
\centering
    
    \includegraphics[width=\textwidth]{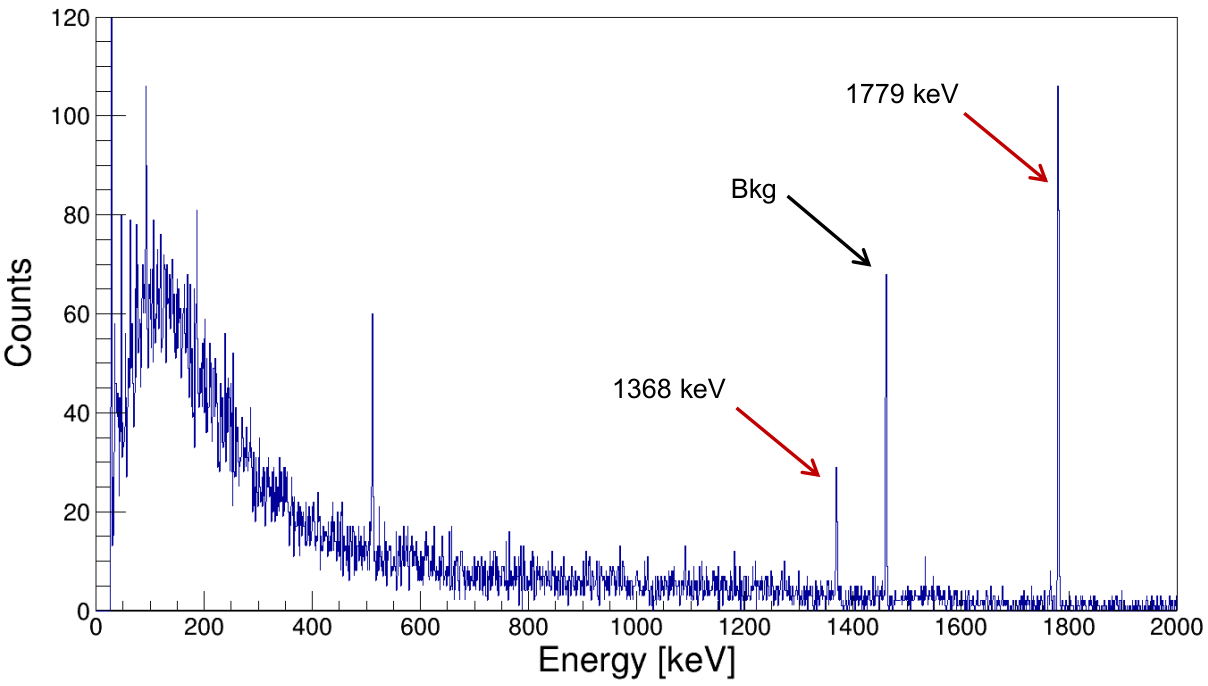}
	\caption{Gamma decay spectrum after irradiation of CeO$_2$ target with $^{13}$C at 10 MeV energy and $\sim$1 p$\mu$A beam intensity. Red arrows indicate the peaks of interest. The black arrow indicates a background peak.}
	\label{fig:ifin-CeO2_T7-Spec}
    
\end{minipage}
\hspace{2mm} \hspace{2mm} \
\begin{minipage}[t]{5.5cm}
\centering

\includegraphics[width=\textwidth]{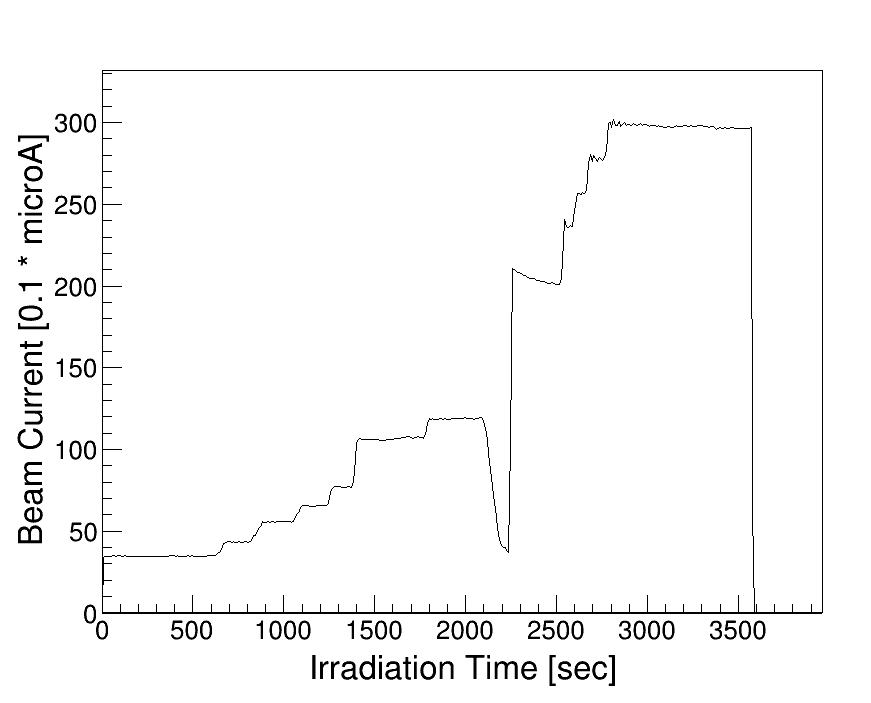}
	\caption{Beam current variation with irradiation time. X-axis shows the time in seconds. Y-axis shows the beam intensity in units of 0.1*$\mu$A.}
	\label{fig:ifin-CeO2_T7-Beam}

\end{minipage}
\end{figure}
This was considered a better option than the previously described $^{13}$C targets and as such the CeO$_2$ pellets were used for the measurement of the $^{13}$C + $^{16}$O cross section, for a range of energies E$_{lab}$ = 8-15 MeV, in steps of 0.4 MeV. Figure \ref{fig:ifin-CeO2_T7-Spec} shows a typical gamma decay spectrum obtained  for this measurement with the target in Figure \ref{fig:ifin-CeO2_T7} at a beam energy of 10 MeV. Two gamma peaks of interest were clearly observed at the expected positions in the spectrum, E$_\gamma$ = 1368 keV from $^{24}$Na and E$_\gamma$ = 1779 keV from $^{28}$Al, as shown by the red arrows. The remaining active channel of the three produced in this reaction was not observed at this energy, as its cross section is significantly lower comparatively. An issue that was noticed immediately was that irradiations could not begin at the highest current available because vacuum would break immediately. Instead, a beam intensity of $\sim$50 pnA was used at the beginning and then increased slowly over the duration of the irradiation (Figure \ref{fig:ifin-CeO2_T7-Beam}). Given the short half-lives of the nuclei of interest, it was necessary for this variation in beam current to be recorded and corrected for in the final yield calculations.

Over the duration of the experiment, it was determined that the targets could be irradiated a maximum of 3 times at the highest currents before becoming unusable and up to 5 times at intensities on the order of $\sim$250 pnA. 
We suspect that the damage sustained by the pellets was due to the heat produced because of the high beam intensity, even though the target holder was equipped with a water-cooling system. 
%We have also found that if the targets are held in vacuum for at least a day before irradiation, they perform better. Finally, it appears 
It was concluded that the best option for such targets is for them to be used for a single measurement/irradiation, upon the availability of the material. Unfortunately, for the study of $^{13}$C + $^{16}$O, the limitation on irradiation time led to an inability in studying the longer-lived activation channel ($^{24}$Na, T$_{1/2}$ = 15 hours) at lower energies and cross sections.

\paragraph{(b) Tantalum pentoxide (Ta$_2$O$_5$)}

\begin{figure}[ht]
\begin{minipage}[t]{5cm}
\centering
	\includegraphics[width=\textwidth]{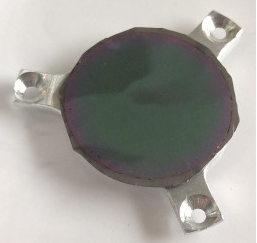}
	\caption{Photograph of the Ta$_2$O$_5$ target with layer thickness of $\sim$438 nm.}
	\label{fig:ifin-Ta2}
\end{minipage}
\ \hspace{2mm} \hspace{2mm} \
\begin{minipage}[t]{7.5cm}
	\centering
	\includegraphics[width=\textwidth]{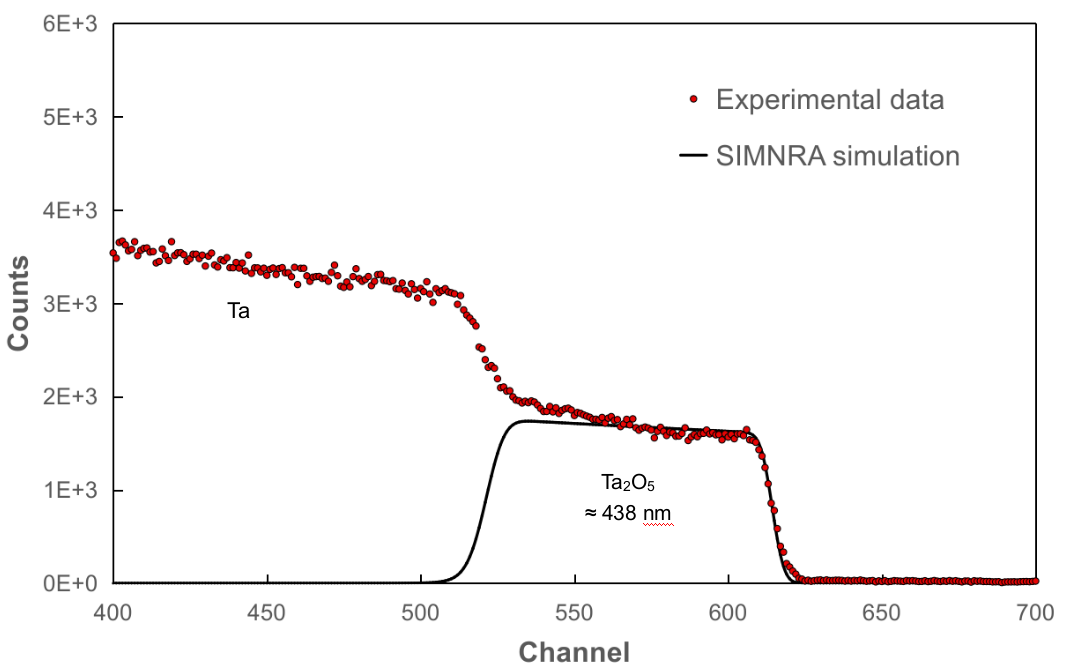}
	\caption{RBS spectrum from the central spot of the target in Figure \ref{fig:ifin-Ta2}. The simulated tantalum oxide layer thickness (black line) does not agree perfectly with the experimental data (red dots), presumably due to the presence of rugosities on the surface (as can also be seen in Figure \ref{fig:ifin-Ta2}).} 
	\label{fig:ifin-Ta2rbs}
\end{minipage}
\end{figure}

Targets of Ta$_2$O$_5$ prepared at Laboratori Nazionali del Gran Sasso of INFN, Italy 
%in collaboration with LUNA 
, by anodization of tantalum backings in ultrapure water, were also used and tested. The details of this technique can be found in the work reported in \citet{Caciolli}. Two targets with two different thickness of the Ta$_2$O$_5$ layer ($\sim$171 nm and $\sim$438 nm) were prepared, the latter is 
%at the  for the purpose of tests with the activation method. One such target is 
shown in Figure \ref{fig:ifin-Ta2}.

Before the activation test, both targets underwent RBS measurements to verify stoichiometry and determine the thickness of the Ta$_2$O$_5$ layer as well as its uniformity. The measurements were carried out at IFIN-HH with the 3 MV Tandetron using a $^4$He beam ($\sim$1 mm$^2$ spot size) and E$_{lab}$ = 2.00 and 3.01 MeV. An example of the scattered $\alpha$ spectra obtained with a silicon detector at 165$^\circ$ can be seen in Figure \ref{fig:ifin-Ta2rbs}, showing the data (red circles) for the central spot of the target in Figure \ref{fig:ifin-Ta2}, as well as the simulation performed with SIMNRA 7.0 (black line). The continuous 'background' indicates the Ta backing, whereas the 'shoulder' on the right shows the Ta$_2$O$_5$ layer.
Both targets were found to have a uniform layer of Ta$_2$O$_5$ and the expected stoichiometry of $\sim$0.4. However, due to the thinner backing, the target in Figure \ref{fig:ifin-Ta2} experienced rugosities in transport, which can be seen in the slight disagreement between experimental data and simulation. The layer thickness for each target was found to be $\sim$171 nm and $\sim$438 nm. 

\begin{figure}[ht]
\begin{minipage}[t]{6.5cm}
\centering
	\includegraphics[width=\textwidth]{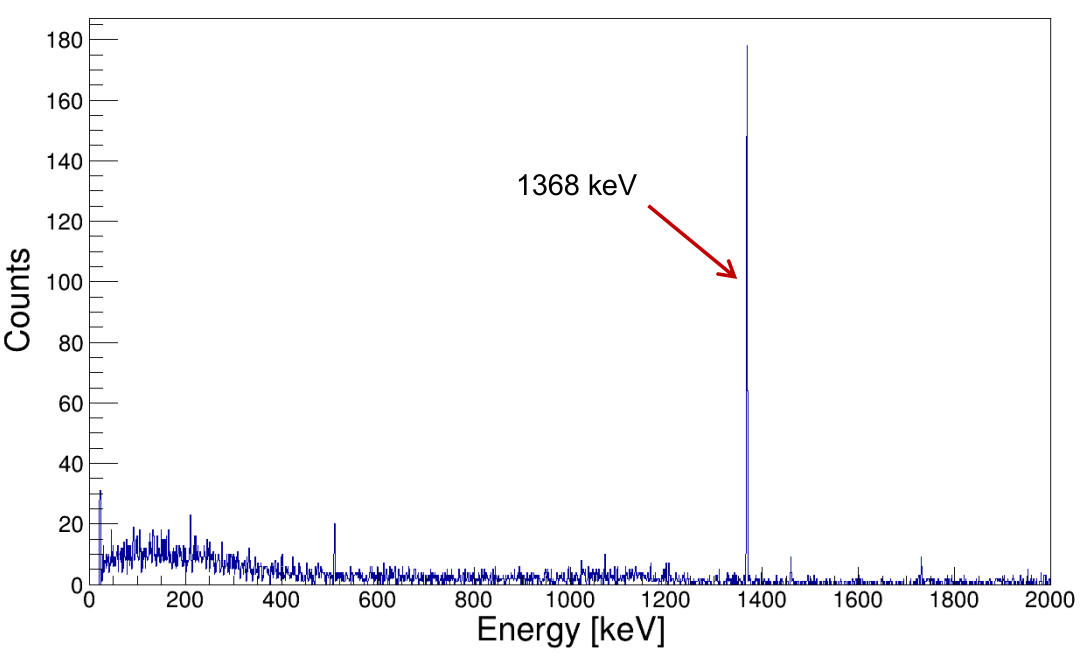}
	\caption{Gamma decay spectrum after irradiation with $^{13}$C at 8 MeV, recorded for $\sim$3.5 hours. The red arrow indicates the peak from $^{24}$Na, at 1368 keV.}
	\label{fig:ifin-Ta2spec}
\end{minipage}
\ \hspace{2mm} \hspace{2mm} \
\begin{minipage}[t]{6.5cm}
	\centering
	\includegraphics[width=\textwidth]{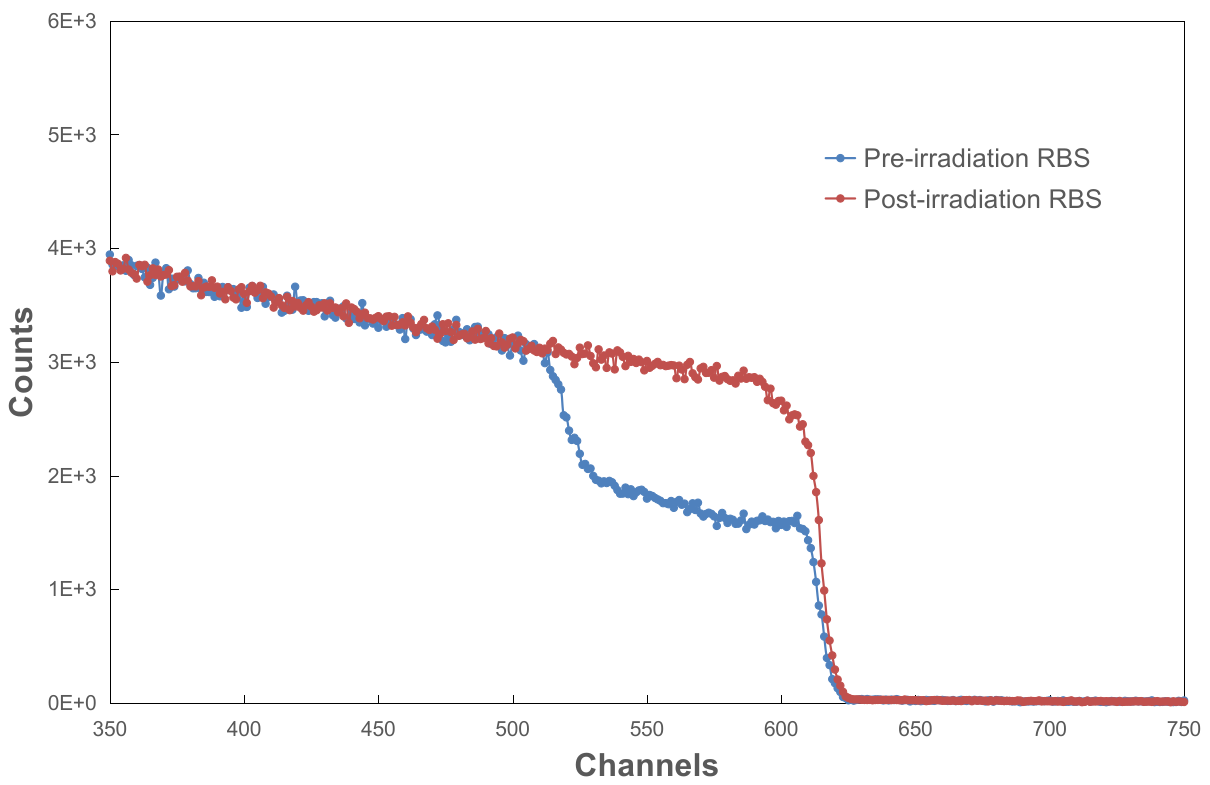}
	\caption{Comparison RBS spectrum from the center of the beam-spot showing the complete degradation of the Ta$_2$O$_5$ layer. Blue line indicates pre-irradiation data. Red line indicates post-irradiation data.}
	\label{fig:ifin-Ta2rbs2}
\end{minipage}
\end{figure}

The activation test was done with a $^{13}$C beam and a focus on the longer lived $^{24}$Na + p$\alpha$ channel (T$_{1/2}$[$^{24}$Na] = 15 h). The goal was to determine the sturdiness of the targets, specifically the oxide layer, under high beam currents and longer irradiations (relative to the previous targets tested). Due to the limited amount of beamtime available, it was not possible to actually test 24-hour long irradiations, which would be necessary at the energies of interest (below E$_{lab}$=7 MeV). 
The targets were irradiated with beams at different energies and with different total beam charge (approximately 30 mC and 70 mC, respectively). Figure \ref{fig:ifin-Ta2spec} shows an example of a gamma decay spectrum obtained from the 438 nm thick Ta$_2$O$_5$ target. The gamma peak marked in red is at E$_\gamma$ = 1368 keV and results from the decay of $^{24}$Na, one of the active channel of interest (-p) for the fusion cross section determination.
%The target was irradiated at 10 MeV with a beam intensity of 7 $\mu$A, on average, for 2 hours, with 30 mC. The second target was irradiated at 8 MeV with a beam intensity of 10 $\mu$A, on average, for 3.5 hours, with 70 mC. The decay spectra for each measurement can be seen in Figures \ref{fig:ifin-Ta1spec} and \ref{fig:ifin-Ta2spec}.\\

%We found that the statistics obtained for these peaks at 8 MeV is a factor of 10 better than that obtained with previous types of targets . 
The targets held well under vacuum and were easy to handle, which was an improvement over the CeO$_2$ pellets that were previously used. After decay measurements, the RBS test was repeated to determine how much deterioration the Ta$_2$O$_5$ layer suffered due to the irradiation. Figure \ref{fig:ifin-Ta2rbs2} shows a comparison between the $\alpha$ spectra pre and post irradiation for the thicker target. The x-axis has been zoomed around the Ta$_2$O$_5$ ‘shoulder’ to emphasize the deterioration. Unfortunately, as it can be seen from the figure, the deterioration was complete in the beam spot region. Furthermore, we cannot estimate at which point during the irradiation the Ta$_2$O$_5$ layer disappears, thus we could not correctly determine any of the quantities of astrophysical interest, such as yield or cross section. The difference is that for this ion irradiation case, the energy loss in the oxide target is much higher than that of the proton reactions studied at LUNA.

\begin{figure}[!ht]
%\begin{minipage}[t]{6.5cm}
\centering
%\begin{center}
   \includegraphics[width=0.8\textwidth]{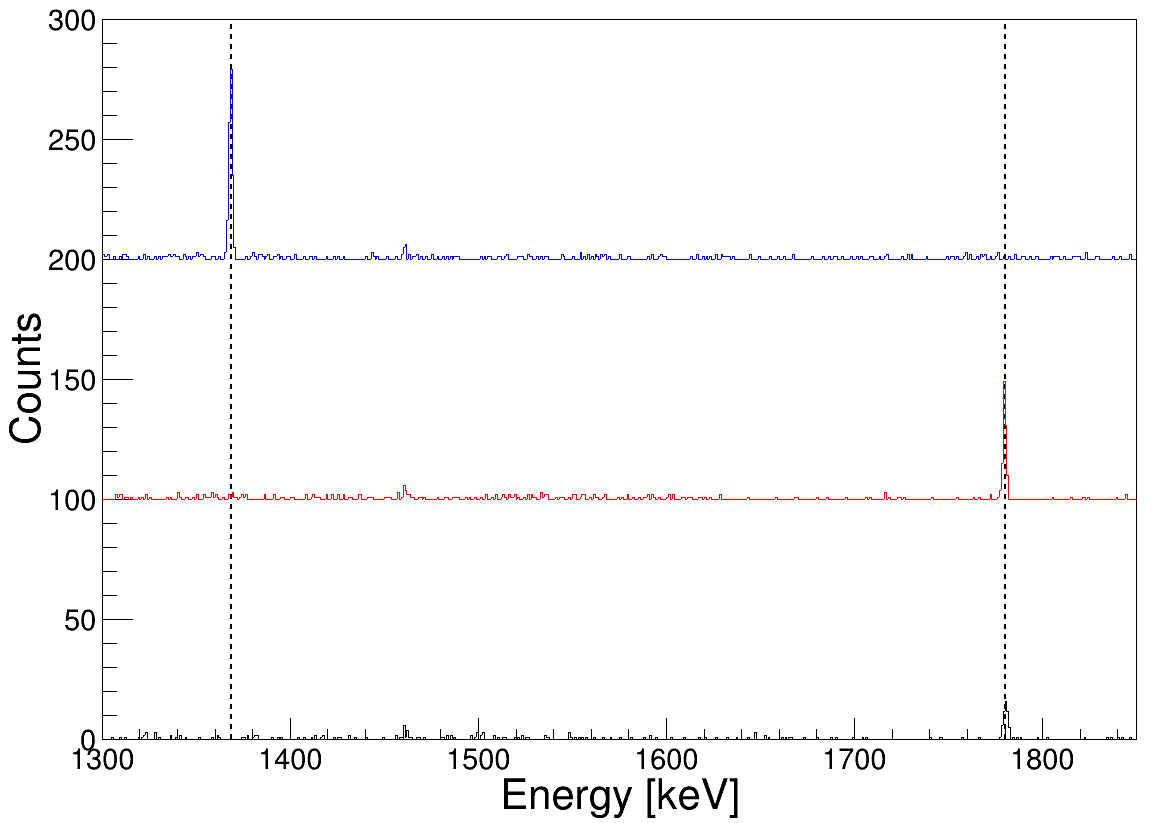}
	\caption{Comparison between gamma decay spectra obtained with the three types of targets described. Blue (top) indicates Ta$_2$O$_5$, red (middle) shows CeO$_2$ and black (bottom) is $^{13}$C. X-axis is zoomed around the two energies of interest which are shown with dashed lines. The spectra have been offset on the vertical axis for easier visibility. All irradiations were done at similar center of mass energies and the spectra were recorded for 5 minutes each.}
	\label{fig:13c16o_tgt_specs}
%\end{center}
\end{figure}

To conclude, Figure \ref{fig:13c16o_tgt_specs} shows a comparison between gamma decay spectra obtained with the three types of targets described in this subsection. All irradiations were done at similar center-of-mass energies and the spectra were recorded for 5 minutes each. The blue line indicates data obtained with a Ta$_2$O$_5$ target, red shows data from a CeO$_2$ pellet and black represents a $^{13}$C target. Two energies are indicated with dashed lines, corresponding to gammas from 2 activation channels: 1368 keV from $^{24}$Na, and the 1779 keV peak from $^{28}$Al. The red peak at 1779 keV in Figure \ref{fig:13c16o_tgt_specs} suggests that, although difficult to handle, CeO$_2$ pellets can be used with currents lower than 1.5 p$\mu$A and irradiation times smaller than 1.5 h. This means they do not allow for the study of the longer-lived activation channel but can be used successfully to measure short-lived nuclei like $^{28}$Al.  
$^{13}$C and Ta$_2$O$_5$ are much better in terms of handling but require a method to monitor the target deterioration during irradiation, as well as post-measurement RBS in order to account for the loss of material. Lastly, in terms of amount of target atoms provided, the anodized Ta$_2$O$_5$ targets appear to contain the most among the three, therefore providing the best statistics. This can be seen in the presence of the blue peak in Figure \ref{fig:13c16o_tgt_specs} from $^{24}$Na and the absence of any overlapping red ones in the CeO$_2$. 
%However, a much more in-depth study is necessary to determine the appropriate combination of layer thickness, backing thickness, beam intensity and collected charge necessary in order for these targets to be used in the astrophysical study of the reaction $^{13}$C + $^{16}$O.\\

\begin{figure}[ht]
%\begin{minipage}[t]{6.5cm}
\centering
   \includegraphics[width=0.9\textwidth]{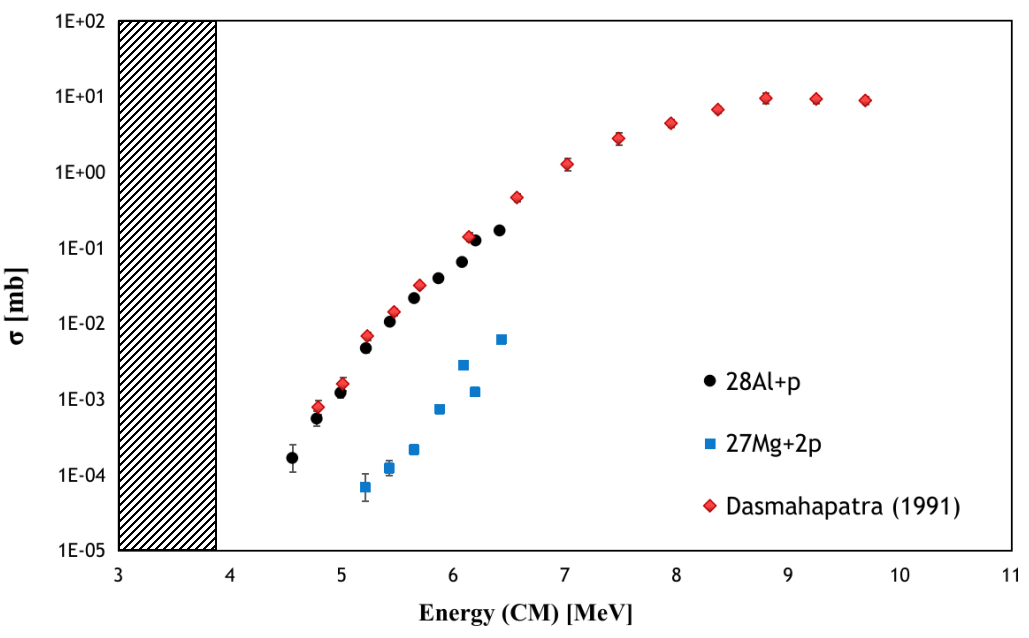}
	\caption{Experimental cross sections as a function of center of mass energy for the two short-lived active channels measured, $^{13}$C+$^{16}$O $\rightarrow$ $^{28}$Al+p and $^{13}$C+$^{16}$O $\rightarrow$ $^{27}$Mg+2p. Error bars indicate statistical errors and are too small to be seen on the plot at higher energies. Red points indicate result from previous literature, Ref \citet{Dasmahapatra}, for this reaction and respective channels. The hatched region shows the Gamow window for this reaction, estimated at T$_9$ = 1K.}
	\label{fig:13c16o_cs}
%\end{center}
%\end{minipage}
\end{figure}

These tests were crucial to decide for the CeO$_2$ targets in the study the $^{13}$C+$^{16}$O reaction with the thick target activation method at IFIN-HH. Figure \ref{fig:13c16o_cs} shows partial results for the experimental cross sections for the two short-lived activation channels that could be measured, $^{28}$Al+p (black points) and $^{27}$Mg+2p (blue points). Comparison with previous studies is also shown with red points (from Ref. \citet{Dasmahapatra}), but only for the p-channel as they did not have measurements for the second one. Despite the limitations present in using the pressed cerium oxide pellets, it was possible to extend the known cross sections towards lower energies% \citet{AlexTBA}
. Further measurements are planned at IFIN to extend this range towards lower energy, in order to support the theoretical studies of fusion mechanisms.
%overlap decay spectra from the three different targets
%\end{minipage}
%\ \hspace{2mm} \hspace{2mm} \
% Add cross section or S-factor here?

\subsection{$^{12,13}$C+$^{19}$F} \label{13c19f}

As part of the ion-ion fusion reactions program, the NAG group at IFIN-HH also intends to study the fusion reactions $^{12,13}$C + $^{19}$F at energies below the Coulomb barrier. 
They fall in the mass region considered "light-light" for fusion systems and %as such a study of the cross section at sub-Coulomb energies could provide useful information regarding the fusion mechanisms that may be occurring there, as well as help answer the current debate on the presence of hindrance in such systems. 
despite their potential for fusion mechanisms studies, or to resolve the matter of the proposed hindrance mechanism, these reactions have not been studied in depth before. 
The most recent study is presented in \citet{Gaedke} and provides yields and cross section results for the two main activation channels that are produced in the reaction $^{nat}$C + $^{19}$F, obtained for the purpose of estimating possible contamination in their other data of interest. The energy range only reaches as low as E$_{CM}$ = $\sim$7 MeV and it is not low enough to determine the presence or absence of the hindrance phenomenon, although it falls under the Coulomb barrier for this system.

Given that $^{13}$C targets have already been tested, as shown in the previous section, the focus of the tests for this reaction was on determining the suitability of two types of fluoride materials: (a) LiF and (b) BaF$_2$. Unfortunately, due to time constraints on the beam availability it was not possible to perform extensive testing, as above.\\

\paragraph{(a) Lithium Fluoride (LiF)}

The first compound that was assessed was lithium fluoride. Targets were made from LiF powder pressed into pellets at 10 Ton. The diameter of the pellets was 1.3 cm and they were also placed on thin aluminum disks for easier handling. One issue that was immediately apparent was that due to the hygroscopic nature of the material, they outgassed more strongly and it took longer to reach the intended beam current during irradiation, even for relatively low intensities. A second consequence of that was the presence of contaminant reactions on hydrogen and oxygen. In particular, reactions on oxygen produce similar gamma peaks to those of interest from fluorine.

\begin{figure}[ht]
\begin{minipage}[b]{6.5cm}
\centering
        \includegraphics[width=\textwidth]{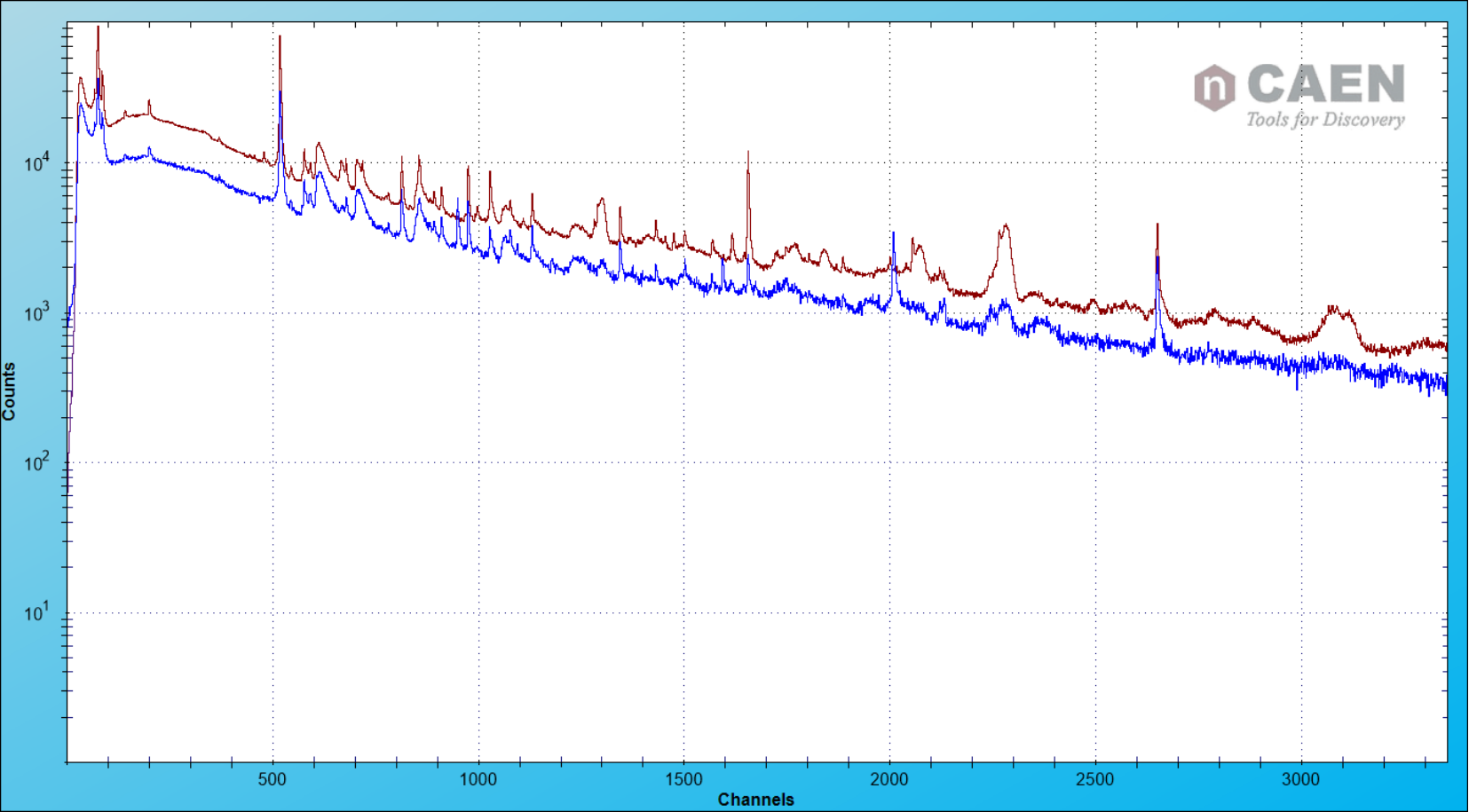}
	\caption{Prompt gamma spectra comparison between the two types of targets. Blue indicates LiF, while red indicates BaF$_2$. Both spectra were corrected for dead time and normalized to beam charge.}
	\label{fig:19f-on}
\end{minipage}
\ \hspace{2mm} \hspace{2mm} \
\begin{minipage}[b]{6.5cm}
	\centering
	\includegraphics[width=\textwidth]{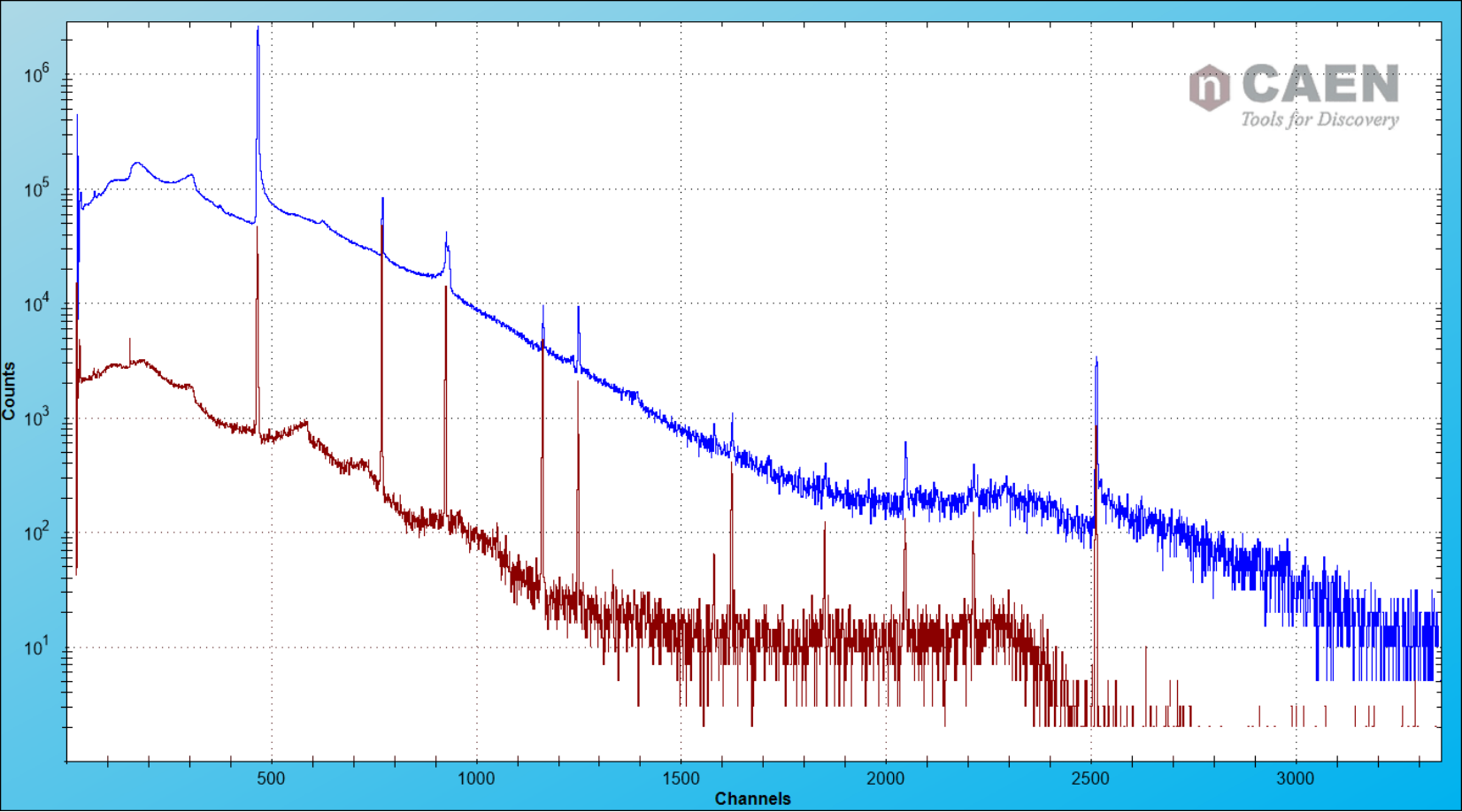}
	\caption{Decay gamma spectra comparison between the two types of targets. Blue indicates LiF, while red indicates BaF$_2$. Both spectra were recorded in 10 minutes and corrected for dead time. }
	\label{fig:19f-off}
\end{minipage}
\end{figure}

A LiF target was irradiated for 1.5 h with a $^{13}$C beam at laboratory energy 18 MeV and intensity $\sim$50 pnA ($\sim$1.3 mC). A visual assessment with an 8x microscope showed a highly visible beam spot but no surface deterioration. The pellet also handled transportation from the beamline setup to the decay station without suffering damage. 
However, the deactivation spectra showed large background from decaying nuclides produced in the reaction on lithium (Figure \ref{fig:19f-off}, blue line). For the purpose of measuring cross sections at low energies where the activation method is more sensitive than prompt measurements, having an increased background is not desirable and therefore, this type of target was found to be unsuited for these experiments.
%it was decided that the Li component created too much contamination in the prompt spectrum and it overshadowed the channels of interest as well as endangered the gamma detectors.\\

\paragraph{(b) Barium Fluoride (BaF$_2$)} 

The second compound that was tested was barium fluoride. The targets were made from BaF$_2$ powder, also pressed into pellets at 10 Ton pressure. Barium has a much higher proton number and as such, the issue of contamination in the spectra was eliminated. The diameter of the pellets was also 1.3 cm and they were placed on rectangular aluminum backings for easier handling.
The main issue here was that BaF$_2$ powder does not compact well and the pellets were more fragile to handle than the LiF ones, so that some of them broke just handling. 
%. Two targets broke just from handling, one before irradiation and one after irradiation (during transport to the decay station). 
It is possible that a larger diameter for the pellets (which would allow a maximum pressure of 25 Ton to be used) will make them sturdier and it is an option to be tested in the future. A BaF$_2$ target was irradiated with the same $^{13}$C beam and under similar parameters for consistency. 

Figure \ref{fig:19f-on} shows a comparison between the prompt gamma spectra obtained with the two types of targets, the blue line representing LiF and red showing BaF$_2$. The most significant difference between them comes from the better statistics due to the increased stoichiometry of fluorine in BaF$_2$. Similarly, Figure \ref{fig:19f-off} shows a comparison between the decay gamma spectra from the two targets (blue shows LiF, red shows BaF$_2$). As mentioned, with BaF$_2$ the peak-to-background ratio is much better compared to LiF, where lithium produces strong contaminant reactions. In conclusion, BaF$_2$ is the preferred option between the two and can be successfully used for the future measurements of $^{12.13}$C+$^{19}$F that do not require beam currents higher than 0.5-1 p$\mu$A. 
%For measurements that do require such high currents, previous experience with pressed pellets indicates that a different type of target must be used.

%figures1314

%\subsection{University Kore and LNS-INFN} \label{lns}
%, who is working for both the STAR goals (sec. \ref{intro}). 

%At the moment, the noble elements solid target facility is under construction, and in the following the results achieved until now are described: 

\section{Highly enriched fluorine target for physics beyond the standard model} \label{beyond}

Highly enriched fluorine targets have also been tried at Physical Chemistry Techniques Department of Laboratori Nazionali del Sud (LNS) of INFN in Catania, Italy.

These targets will be used also for the NEW JEDI experiment \citet{Beyhan}, which aims to explore potential physics beyond the standard model and in particular is devoted to the X17 boson search through the study of $^{8}$Be level decay, populated via $^7$Li+p, where LiF targets are being used. 
As usual when the desired element (lithium) is not the only one in the target, to retrieve its contribution, highly enriched targets of the other elements present in the target (fluorine) are needed to determine the background produced. 

\paragraph{CaF$_2$ targets}
CaF$_2$ was chosen as a first attempt to produce targets with high fluorine content (usually in the form of salt). The PVD (Physical Vapor Deposition) deposition technique was used as a vacuum evaporation technique. 
The salt was deposited on amorphous graphite thin layers, 30 $\mu$g/cm$^2$ thick. It is worth noticing here that heating the sample to 180$^{\circ}$ was necessary before the salt evaporation, to remove the absorbed water by this very hygroscopic salt. Therefore, in high vacuum conditions, the salt placed inside the evaporator sublimates when heated. To control the evaporation process, a quartz microbalance was used, measuring the speed of the deposited salt as a function of time.
To have further feedback of the target thickness, the $\alpha$-particles scanning system was used, returning the value of 150 $\mu$g/cm$^2$. 

\paragraph{PTFE targets}
To overcome the important issues related to the salt hygroscopicity, looking for an easy to use and handle target, among the other investigated materials, a good choice has turned to be polytetrafluoroethylene PTFE, $-$(F$_2$C$-$CF$_2$)n$-$, also known as teflon. It is a plastic polymer with a large amount of fluorine ($\frac{C}{F}$=$\frac{2}{4}$) and density of 2.2 g/cm$^3$. 
The technique chosen was cold lamination. Even though teflon is a plastic material, at each lamination step the two rollers distance was reduced of 1/6 of a millimeter, to thin it optimally.
The thickness of three teflon targets obtained was measured with the automatic $\alpha$ scanning machine, measuring the energy loss in the film of the $\alpha$ particles coming from $^{241}$Am decay.
Results are listed in Table \ref{tabteflon}, with the mean value obtained and their standard deviation (STD). 

\begin{threeparttable}[ht]
\caption{Characterization of the new teflon target produced.\label{tabteflon}}
%	\begin{adjustwidth}{-\extralength}{0cm}
		%\newcolumntype{C}{>{\centering\arraybackslash}X}
	\centering	\begin{tabular}{|c|c|c|c|}
			\toprule
			\textbf{sample}	& \textbf{Measure}	& \textbf{$\Delta$E [keV]}     & \textbf{Thickness [$\mu$g/cm$^2$]} \\
			\midrule
\multirow[m]{3}{*}{DP490}  & 1 & 708.47 & 1051 \\
			  	    & 2	& 724.02 & 1074\\
			              & 3 & 714.80 & 1061\\
                        & mean & 715.76 & 1062 \\
                        & STD & 7.82  &   12    \\
                   \midrule
\multirow[m]{3}{*}{DP491}   & 1 & 850.37 & 1262 \\
			  	    & 2	& 852.57 & 1265 \\
			              & 3 & 843.90 & 1252 \\
                 & mean &848.95 &  1260 \\
                 & STD &  4.51  &  7         \\
                   \midrule
\multirow[m]{3}{*}{DP492}   & 1 & 786.19 & 1167 \\
			  	    & 2	& 792.34 & 1176 \\
			              & 3 & 786.34 & 1167 \\
                 & mean & 788.29 &  1170 \\
                 & STD & 3.51   &    5  \\
                 
			\bottomrule
		\end{tabular}
	\end{threeparttable}

\begin{figure}[ht]
\begin{center}
\centering
        \includegraphics[width=0.3\textwidth]{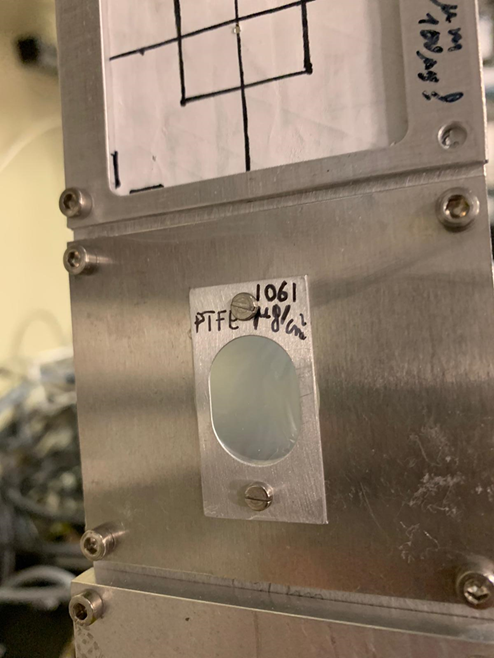}
	\caption{PTFE target tested with a Cyclotron proton beam of NPI of UJF.}
	\label{teflon}
\end{center}
\end{figure}

During the NEW JEDI experiment commissioning at UJF Nuclear Physics Institute (NPI) at Řež (Prague - Czeck Republic), those targets were tested under proton beam (Figure \ref{teflon}), where they proved to resist up to 200 nA of beam current, then underwent a quick deterioration.

\begin{figure}[ht]
\begin{center}
\centering
        \includegraphics[width=0.6\textwidth]{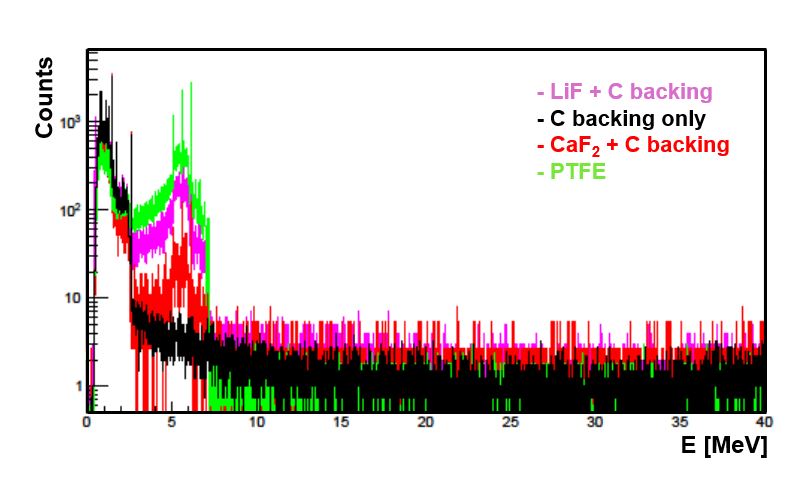}
	\caption{Yields from one detector of the NEW JEDI experiment obtained by a 1.070 MeV proton beam on different targets. Pink line is for a 154 $\mu$g/cm$^{2}$ LiF target with a 40 $\mu$g/cm$^{2}$ C backing. Black line for a 40$\mu$g/cm$^{2}$ C target. Red line for a 153 $\mu$g/cm$^{2}$ CaF$_2$ target with a 34 $\mu$g/cm$^{2}$ C backing. Green line for a 1061 $\mu$g/cm$^{2}$ PTFE target.}
	\label{nj}
\end{center}
\end{figure}

In the commissioning period of NEW JEDI experiment, also the fluorine targets were tested. In Figure \ref{nj} the yields registered by one of the detectors are shown, normalized to one peak located at 2.6 MeV. These test were made with a proton beam of 1.070 MeV on 154 $\mu$g/cm$^{2}$ LiF target with a 40 $\mu$g/cm$^{2}$ C backing (pink line), a 40$\mu$g/cm$^{2}$ C target (black line), a 153 $\mu$g/cm$^{2}$ CaF$_2$ target with a 34 $\mu$g/cm$^{2}$ C backing (red line) and 1061 $\mu$g/cm$^{2}$ PTFE target (green line). Results from carbon and fluorine targets will be used for the background subtraction.

\section{Nucleosynthesis via s-process and $\alpha$-induced reactions} \label{s-proc}

In recent times, nuclear astrophysicists have directed their attention toward cross sections pertinent to the s-process, taking into account the possibility that r-process abundances can be inferred from those of the s-process. The s-process nucleosynthesis follows a trajectory that meanders along the valley of $\beta$-stability, in proximity to the strongly bound isotopes of a given mass number A. The primary focus for slow neutron captures lies in the final Asymptotic Giant Branch (AGB) phase of low- and intermediate-mass stars \citet{neutrondriven}. Stars with a mass up to 3 M$_{\odot}$ are, in fact, responsible for generating the main component of the s-process (i.e., nuclei ranging from Sr to Bi). The primary neutron source is known to be the $^{13}$C($\alpha$,n)$^{16}$O reaction \citet{BUS99}. Operating at typical energies of 8 keV under radiative conditions, this reaction provides a neutron density of about 10$^{6\div7}$ n/cm$^3$ \citet{GAL88} \citet{Ili10}. 
%\textbf{The $^{22}$Ne($\alpha$,n)$^{25}$Mg reaction serves as the second neutron source in AGB stars, being marginally activated only during the thermal pulses (periodic episodes of He-convective burning) in low-mass stars. Only red giants more massive than 3 M$_{\odot}$ (intermediate-mass stars) efficiently burn $^{22}$Ne, reaching maximum temperatures of $3 - 3.2 \cdot 10^8$ K. Even stars with larger masses, concluding their evolution as core-collapse supernovae, contribute to s-process nucleosynthesis, particularly to the weak component responsible for the production of nuclei with A$\leq$85 \citet{THE07}. }

A pivotal role in the s-process is played by the $^{22}$Ne($\alpha$,n)$^{25}$Mg reaction, the primary neutron producer for its weak component (60$<$A$<$90) in massive stars, influencing all the late stages of stellar evolution until their demise. Despite the significance of this reaction, three orders of magnitude discrepancies persist in the cross section data within the energy range relevant for astrophysics (E$_{cm}$=300-900 keV), rendering such data essentially unusable for astrophysical purposes \citet{Longland, ads21, Wiescher2023}.

The trend of the excitation function measured so far indicates a steep drop in the yield, allowing for the provision of only upper limits to the cross section, even at center-of-mass energies around 800 keV, which is at the edge of the region of interest. A comprehensive overview of the current state of the field can be found in \citet{Jaeger}. Direct measurements at such low energies are extremely challenging due to the exponential Coulomb damping (the Coulomb barrier is located at 3.5 MeV) of the cross section, reducing values to less than 1 $\mu$b and essentially pushing the signal-to-noise ratio to zero.

For this reasons, LUNA collaboration is working on a direct measurement of both the $^{22}$Ne($\alpha$,n)$^{25}$Mg and the competitor $^{22}$Ne($\alpha$,$\gamma$)$^{26}$Mg reaction cross sections at the B-IBF, exploiting the high intensity beam of LUNA-MV accelerator \citet{Sen2019}, the ultra-low background, an extended $^{22}$Ne gas target and a properly designed detector array \citet{Ananna2024b, Ananna2022, Mercogliano2024}.
Nonetheless, indirect measurements, such as capture and transfer reactions \citet{Jaeger}, are necessary at low energy to provide resonance parameters, such as spectroscopic factors, for use in the calculation of the reaction rate. In this context, a recent paper \citet{ads21} recalculated the reaction rates of $^{22}$Ne($\alpha$,n)$^{25}$Mg and its competitor $^{22}$Ne($\alpha$,$\gamma$)$^{26}$Mg, updating energies and spin-parities of all the $^{26}$Mg levels, with a particular focus on the results in \citet{Hesh}, obtained through $^{22}$Ne($^{6}$Li,d)$^{26}$Mg and $^{22}$Ne($^{7}$Li,t)$^{26}$Mg transfer reactions. 
%The authors find that, excluding the results in \citet{Hesh}, s-process nucleosynthesis will not substantially change. However, to 
This leads to a better agreement of Ba and Zr isotopic ratios with data from presolar SiC grains. Nevertheless, the results reported in \citet{ads21} and \citet{Hesh} suggest the need for further measurements, particularly focusing on the excitation function trend, to also account for levels interference in the evaluation of the reaction rate.

\begin{figure}[ht]
\begin{center}
\centering
        \includegraphics[width=0.4\textwidth]{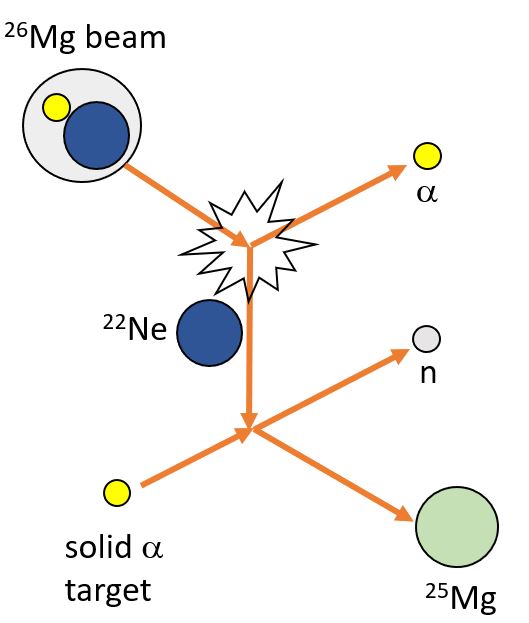}
	\caption{Sketch of a possible reaction to measure to get the $^{22}$Ne($\alpha$,n)$^{25}$Mg cross section with the THM, taking advantage of the solid helium targets to be produced at LNS-INFN.}
	\label{sketch}
\end{center}
\end{figure}

To deal with this, a possible solution is a new measurement of the $^{22}$Ne($\alpha$,n)$^{25}$Mg with the indirect Trojan Horse Method (THM) \citet{Spit19, Tumino21,Tumino2025}, using a $^{26}$Mg beam as a $^{22}$Ne inducer, the two-body reaction of interest being half of the energy shell. A sketch of the three-body reaction to measure to get the $^{22}$Ne($\alpha$,n)$^{25}$Mg cross section at very low energies, is represented in Figure \ref{sketch}. THM results will be the better the lower the energy reached by LUNA measurement, because of the normalization needed by the method itself. 

The use of solid $\alpha$-targets is conditional to keep the low energy and angular spread needed by the THM to be used. Thus, noble elements solid targets are the next years challenge, that will be attempt at LNS of INFN in Catania, coupling the local Tandem pre-acceleration system and an ECR source who can provide also noble elements, with the idea to implant He and Ne into a host material, being impossible for them to create solid compounds, and then to test them to follow their stability under beam bombardment. 
 
%%%%%%%%%%%%%%%%%%%%%%%%%%%%%%%%%%%%%%%%%%
\section*{Summary and perspectives}

Substantial progresses in target preparation protocols at the STAR members laboratories are presented, in the framework of the ChETEC-INFRA project.  In particular, pure and stable $^{19}$F targets to match new experimental needs of nuclear astrophysics experiments, came out from investigations on new materials and protocols carried out both at IFIN-HH and LNS-INFN. At IFIN-HH also new $^{16}$O and $^{13}$C targets have been produced and tested for the study of the hindrance effect in light ion fusion at sub-barrier energies. 
Results shown claim for continuing the testing under beam bombardment of fluorine targets at IFIN-HH and the search for new and more resistant compounds for oxygen targets.
Moreover, huge efforts have been dedicated in the research of $^{12}$C targets to improve experimental chances to measure the fusion $^{12}C+^{12}C$ reactions, critical to understand massive star evolution, as done by the STELLA collaboration. Tests of target stability under high intensity beam irradiation at Felsenkeller and H and D desorption from $^{12}$C targets of the HEAT project, aim at suitable targets for the $^{12}C+^{12}C$ cross section measurement planned by LUNA collaboration. Noble elements solid targets is the next years challenge, that for example will disclose the feasibility of an indirect measurements with THM of the $^{22}$Ne($\alpha$,n)$^{25}$Mg cross section in the relevant energy region for astrophysics and, thus, to fix s-process models weaknesses.

For the first time, European target laboratories have been gathered in a context where they can learn from mutual experiences and cooperate.  However, more than this, the whole European nuclear astrophysics community has shed light on the key role of the targets in the experiments. The problems in measuring cross sections coming from target non-uniformity effect, low resistance to intense beam currents and presence of contaminants, still require to continue these devoted studies by the STAR.

Thus, rather than the single laboratory improvement, the outlook for the future the possibility to improve together to meet the challenges of experimental requirements. This will be crucial for the step forward needed in nuclear astrophysics for the understanding of C-burning and the s-process. 

%%%%%%%%%%%%%%%%%%%%%%%%%%%%%%%%%%%%%%%%%%

\section*{Funding}
This work was partially supported by the European Union ChETEC-INFRA, project no. 101008324. Parts of this work is supported by the Romanian Ministry of Research, Innovation and Digitization under the Nucleus Programme, Contract PN 23 21 01 02. Use of the 3 MV TandetronTM accelerator at IFIN-HH was financially supported by the Romanian Government Programme through the National Programme for Installations of National Interest (IOSIN). J.N. acknowledges support from the Interdisciplinary Thematic Institute QMat, as a part of the ITI 2021–2028 program of the University of Strasbourg, CNRS and Inserm, which was supported by IdEx Unistra (ANR 10 IDEX 0002), and by SFRI STRAT’US project (ANR 20 SFRI 0012) and EUOQMAT ANR-17-EURE-0024 under the framework of the French Investments for the Future Program. S.C. acknowledges support from the Marguerite Perey Chair of the University of Strasbourg Institute for Advanced Studies.

%%%%%%%%%%%%%%%%%%%%%%%%%%%%%%%%%%%%%%%%%%
%\bibliographystyle{unsrtnat}
\bibliography{biblio}

\end{document}